\documentclass[letterpaper, twocolumn, superscriptaddress, showpacs, preprintnumbers, amsmath, amssymb,longbibliography, nofootinbib]{revtex4-1}
\usepackage[pdftex]{graphicx}
\usepackage{caption}
\usepackage{subcaption}
\usepackage{dcolumn}
\usepackage{bm}
\usepackage{color}
\usepackage{hyperref}
\usepackage{adjustbox}
\usepackage{color}
\usepackage{float}
\usepackage{framed}
\usepackage{esvect}
\usepackage{amsfonts,amssymb,amsmath,hyperref,hypcap,enumerate}
\usepackage{wasysym}
\usepackage{slashed}
\usepackage{tikz}
\usepackage{amsmath,amssymb,bm,graphicx,dsfont}
\usepackage[latin9]{inputenc}
\usepackage{amsmath}
\usepackage{slashed}
\usepackage{dsfont}
\usepackage{amstext}
\usepackage{amssymb}
\usepackage{amsbsy}
\usepackage{amsthm}
\usepackage{epsfig}
\usepackage{tikz}
\usepackage{bbm}
\usepackage{hyperref}
\usepackage{color}
\usepackage{multirow}
\usepackage{ulem}
\usepackage{soul}

\newcommand{\be}{\begin{equation}}
\newcommand{\ba}{\begin{align}}
\newcommand{\ee}{\end{equation}}
\newcommand{\bea}{\begin{eqnarray}}
\newcommand{\eea}{\end{eqnarray}}
\newcommand{\beq}{\begin{equation}}
\newcommand{\eeq}{\end{equation}}
\newcommand{\beqn}{\begin{eqnarray}}
\newcommand{\eeqn}{\end{eqnarray}}

\renewcommand{\vec}[1]{{\bf #1}}
\renewcommand{\hat}[1]{{\widehat #1}}

\begin{document}
\title{ Doping the chiral spin liquid - topological superconductor or chiral metal?
}
\author{Xue-Yang Song}
\author{Ashvin Vishwanath}
\author{Ya-Hui Zhang}
\affiliation{Department of Physics,  Harvard University,
Cambridge, MA 02138, USA}
\date{\today}

\begin{abstract}
We point out that there are two different chiral spin liquid states on the triangular lattice and discuss the conducting states that are expected on doping them. These states labeled CSL1 and CSL2  are associated with two distinct topological orders with different edge states, although they both spontaneously break time reversal symmetry and exhibit the same quantized spin Hall conductance. While CSL1 is related to the Kalmeyer-Laughlin state, CSL2 is the $\nu =4$ member of  Kitaev's 16 fold way classification. Both states are  described within the Abrikosov fermion representation of spins, and the effect of doping can be accessed by introducing charged holons. On doping CSL2, condensation of charged holons leads to a topological d+id superconductor. However on doping CSL1, in sharp contrast, two different scenarios can arise:  first, if holons condense, a chiral metal with enlarged unit cell and finite Hall conductivity is obtained. However, in a second  novel scenario, the internal magnetic flux adjusts with doping and  holons form a bosonic integer quantum Hall (BIQH) state. Remarkably, the latter phase is identical to  a  $d+id$ superconductor.  In this case the Mott insulator to superconductor transition is associated with a bosonic variant of the integer quantum Hall plateau transition. We connect the above two scenarios to two recent numerical studies of doped chiral spin liquids on triangular lattice. Our work clarifies the complex relation between topological superconductors, chiral spin liquids and quantum criticality. 
\end{abstract}

\date{\today}

\maketitle

\section{Introduction}

Ever since Anderson conceived  of the resonating-valence-bond(RVB) liquid as a precursor state to the superconductor in the cuprates \cite{anderson_1987}, the notion of quantum spin liquids (QSLs) has been inextricably tied to mechanisms for high-temperature superconductivity. A considerable body of  work has explored doping spin liquid or Mott insulators\cite{rmp_2006, dopesl_wen_1990,dopesl_rokhsar_1993,dopesl_sigrist_1994,dopesl_fabrizio_1996,Senthil_2000,dopesl_senthil_2005,dopesl_konik_2006,dopesl_balents_2007,dopesl_watanabe_2004,dopesl_karakonstantakis_2013,dopesl_chen_2013,dopesl_white_2015,dopesl_venderley_2019,dopesl_li_2015,dopesl_kelly_2016,dopesl_ye_2016,dopesl_jiang_2020,dopesl_peng_2020,dopesl_gannot_2020}.  
In  referring to  QSLs, it is customary to describe them in terms of two features - the emergent gauge group, eg. $Z_2$ and U(1) spin liquids and additionally the statistics of ``spinon" excitations carrying spin 1/2 and gauge charge. Here we will  consider the case of fermionic spinons\cite{fspinon_wen_20021,fspinon_wen_20022}.  A Mott insulator to superconductor transition upon doping is quite natural when  the parent Mott insulator is a $Z_2$ spin liquid\cite{rmp_2006}. Note, a $Z_2$ spin liquid can be described by a parton mean field ansatz with fermionic spinons in a BCS state. Doped charges are accounted for in terms of bosonic holons, which can condense. Holon condensation makes the pairing of spinons evolve into true Cooper pairs of electrons, leading to a superconducting phase.   In contrast, doping a $U(1)$ spin liquid with a spinon Fermi surface naturally gives us a Fermi liquid phase, rather than a superconductor, because there is no pairing of spinons in the parent Mott insulator\cite{senthil_2008}. 

This brings us to the subject of this paper, the chiral spin liquid(CSL)\cite{kalmeyer1987equivalence,wen_1989}, and the conducting states that emerges on doping. Chiral spin liquids spontaneously break time reversal symmetry and exhibit chiral edge modes. In particular, a speculative connection between fractional statistics that exists in chiral spin liquids and two-dimensional superconducting states, so-called anyon superconductivity, has been studied  in the earlier days of  high Tc superconductivity\cite{chen_1989,lee_1989, wen_1989, lee_1990, laughlin_1991}.   Theoretically, chiral spin liquids have been found to be the ground state for various spin 1/2 lattice models on the kagome and triangular lattices\cite{kagome_bauer_2014,he_2014,gong_2014,kagome_he_2015, zaletel_2020,hu_2016,wietek2015nature} and also in $SU(N)$ model with $N>2$\cite{Hermele_2011,Nataf_large_N_CSL_2016,Poilblanc_2020,zhang_2020}. On the experimental side, recent observation of a quantized thermal Hall effect in a Kitaev material\cite{kasahara2018majorana} suggests a non-Abelian chiral spin liquid. These examples motivate us to ask  - what is the fate of a chiral spin liquid on doping? Surprisingly we find that this question is more complicated and potentially also more interesting than the usual situation of simple $Z_2$ spin liquid, where the conventional RVB theory works well.

The first complexity is that there are actually two different chiral spin liquids: CSL1 and CSL2. CSL1 is an analog of fractional quantum Hall state in spin system, as first proposed by Kalmeyer-Laughlin\cite{kalmeyer1987equivalence}. 
In contrast, CSL2 should be thought as a ``gauged" chiral superconductor, whose wavefunction is obtained by Gutzwiller projection of a spin singlet superconductor with $d_{x^2-y^2}+id_{xy}$ superconducting pairing, abbreviated as $d+id$ pairing. 
Both CSL1 and CSL2 can be conveniently described using mean field ansatz of Abrikosov fermions, generically not gauge equivalent to each other.  For CSL1, the fermionic spinons are put in a Chern insulator phase with $C=2$. In contrast, the spinon is in a $d+id$ superconductor ansatz in the CSL2, very similar to a $Z_2$ spin liquid in terms of wavefunction. Hence the conventional RVB theory can be generalized to this case and a $d+id$ superconductor naturally emerges from doping.  A dual viewpoint involves beginning in the $d+id$ superconductor, and driving a superconductor-insulator transition by condensing pairs of vortices \cite{Balents_1998}, leading directly to CSL2.

On the other hand, the wavefunction of CSL1 is a Gutzwiller projection of Chern insulator. Thus a superconductor is not expected from the RVB picture.   As far as we know, all of the chiral spin liquids found in numerics for spin rotation invariant spin $1/2$ model are CSL1\cite{kagome_bauer_2014,he_2014,gong_2014,kagome_he_2015, zaletel_2020,hu_2016}.  This strongly suggests CSL1 as a more likely candidate relevant to realistic spin $1/2$ materials.  Then a natural question that arises is: what should we expect on doping this chiral spin liquid?  

 The main result of this paper is to propose two different scenarios for doping CSL1 on the triangular lattice.   In the first scenario, a simple generalization of RVB theory using slave boson condensation\cite{rmp_2006} predicts a doped Chern insulator. On triangular lattice, this is a chiral metal with doubled unit cell and staggered loop current order. It also breaks the $C_6$ rotation symmetry completely and has a non-zero electrical Hall conductivity.  This is well described by a mean field theory of electron, which basically inherits the mean field ansatz of spinons in the Mott insulator.  An unconventional second possibility is a $d+id$ superconductor emerges from a completely different mechanism, which can not be captured by simple mean field theory.  The doped holon forms a bosonic state with integer quantum Hall(bosonic-IQH) effect \cite{lu2012,senthil_bIQHE}. The bosonic holons lead to an opposite Hall conductivity ($\sigma^b_{xy}=-2$) compared to the fermionic spinons.  The final wavefunction of electron is a product of a bosonic IQH state and a fermionic IQH state.  This construction, surprisingly, turns out to describe a superconductor with the same topological property as the $d+id$ superconductor as we will show below.
 
Our proposed two scenarios for doping CSL1 are in agreement with with two recent numerical studies of doped chiral spin liquid on triangular lattice.  Ref.~\onlinecite{jiang_2020} finds a $d+id$ superconductor from doping a CSL in the strong Mott regime described by a $J_1-J_2-J_{\chi}$ model\cite{Gong_2017,Saadatmand_2017}.  In contrast, Ref.~\onlinecite{zhu_2020} observes a chiral metal phase  upon doping a CSL in the weak Mott regime \cite{zaletel_2020}.  The CSLs in the parent Mott insulators in the above two cases are shown to be both the CSL1 because the measured entanglement spectrum is consistent with the $SU(2)_1$ edge theory \cite{Gong_2017,zaletel_2020}.   Lastly we note that there are other numerical studies of possible superconductivity on triangular lattice where the parent state is not a chiral spin liquid\cite{venderley2019density,jiang2019superconductivity,kivelson2020}. Further work will be needed to synthesize these observations  into a coherent theory. However it is clear that there are at least two competing phases upon doping the same CSL, which is consistent  with the framework in our paper. 

\section{Two chiral spin liquids on triangular lattice}
\label{parton}
We adopt a parton construction to decompose the physical electron operator as a boson carrying its electric charge (called `holon`) and a fermionic spinon carrying spin indices:
\begin{equation}
    c_{i,\sigma}=b_i^\dagger f_{i,\sigma}
    \end{equation}
where $\sigma,i$ denote spin and site, respectively. This formalism introduces a $U(1)$ gauge redundancy between the holon and spinon, with a $U(1)$ gauge field $a$. Hence it leads to a gauge constraint $n_i^b+n_i^{f}=1$. Note that there is an $SU(2)$ version of the parton construction with two slave bosons per site\cite{rmp_2006} reviewed in appendix \ref{app:parton}, and for our purpose the above $U(1)$ partons suffice.

At integer filling $n_i=1$, the holons are trivially gapped and the mean-field for spinons is conveniently written in the basis $\psi_i=(f_{i,\uparrow},f_{i,\downarrow}^\dagger)^T$, with Pauli matrices $\tau^{x,y,z}$ acting in such spinor space. 
\begin{eqnarray}
\label{hspinon}
    \mathcal H_{spinon}=\sum_{\langle ij\rangle}\psi_j^\dagger u_{i,j}\psi_i,
    \end{eqnarray}
where one considers only nearest neighbor terms. $u$'s are $2\times 2$ matrices, satisfy $u_{i,j}=u_{j,i}^\dagger$ and contain hopping and pairing as diagonal, off-diagonal elements, respectively. We use coordinates shown at upper right of fig \ref{fig:chiral_metal}.

There are two kinds of chiral spin liquid variational states that break time reversal and mirror symmetry on triangular lattices that have been extensively studied:  the $U(1)_2$ chiral spin liquid (CSL1) and the projected $d+id$ superconductor (CSL2). Below we review the mean field theory for the two states and discuss their connection and differences.


\subsection{CSL1: $U(1)_2$ chiral spin liquid}

\begin{figure}
 \captionsetup{justification=raggedright}
    \centering
     \adjustbox{trim={.08\width} {.0\height} {.1\width} {.0\height},clip}
   { \includegraphics[width=0.6\textwidth]{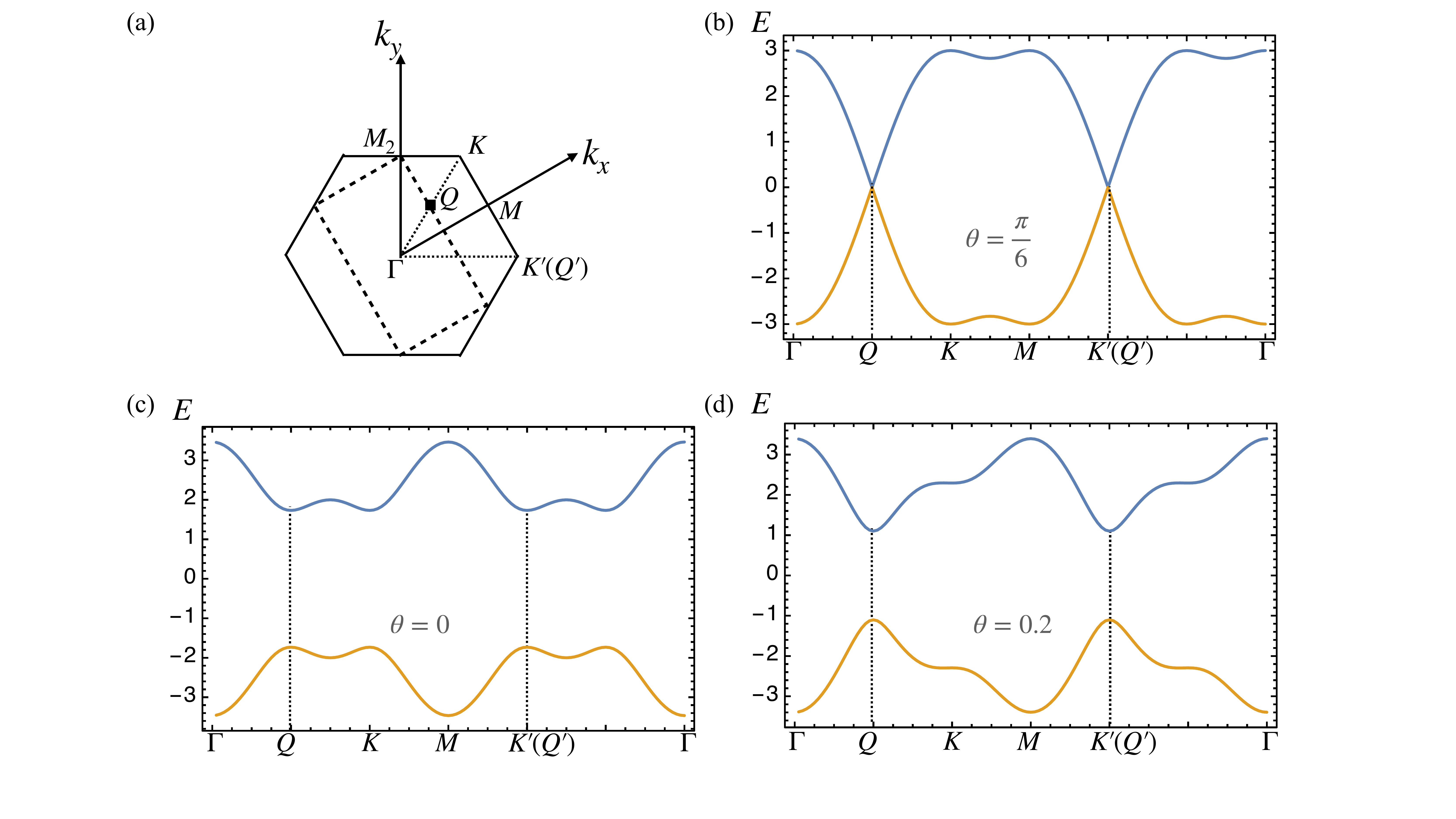}}
 \caption{Dispersion of spinons for $\mathcal H_{spinon}$ in eq \eqref{hspinon}.The band is particle-hole symmetric.(a) The original(reduced) hexagonal(square, dashed) Brillouin zone and the dispersion plots are along the loop $\Gamma\rightarrow K\rightarrow K'\rightarrow \Gamma$ where $\vec Q=(\pi/3,\pi/3),\vec Q'=(4\pi/3,-2\pi/3)$ are points of maximal energy in the valence band. $\Gamma,M_2$ are points of lowest energy in valence band. (b) The dispersion at the DSL gapless point $\theta=\pi/6$. (c)(d) Dispersions at $\theta=0,0.2$, respectively. The Chern number of valence (conduction) bands is $C=\pm 1$.}
    \label{fig:dispersion}
\end{figure}

The $U(1)_2$ chiral spin liquid, which we study in this paper, is described by a Chern insulator of the spinons with total Chern number (combining the two spins) $C=2$. Typical realizations consist of a $\pi$ hopping flux of each spinon $f_\sigma$ through a rhombus of the triangular lattice, with the flux $\pi/2\pm 3\theta$ around an upward and downward triangle for the spinon. This staggered-flux ansatz typically gives a gapped dispersion for each spinon species, forming two Chern bands with $C=\pm 1$; at the point $\theta=\pm \pi/6$, the dispersion becomes gapless and a Dirac spin liquid emerges\cite{DSL_2019}. The mean-field ansatz of spinon hopping reads 
\begin{align}
\label{hspinon}
    &u_{r,r+\hat x}=ie^{i(\frac{\pi}{2}+\theta) \tau^z},\nonumber\\
    &u_{r,r+\hat y}=(-1)^{r_x}ie^{i(\frac{\pi}{2}+\theta)\tau^z},\nonumber\\
    &u_{r,r+\hat x+\hat y}=(-1)^{r_x}ie^{-i\theta \tau^z},(\theta\in (-\pi/3,\pi/3))
    \end{align}
where the $x,y$ coordinates are shown in figure \ref{fig:chiral_metal}. In this gauge choice, the lower spinon band hits the lowest energy at zero momentum and the highest energy at momenta $(\pi/3,\pi/3), (4\pi/3,-2\pi/3)$, independent of $\theta$. See Fig.~\ref{fig:dispersion} for the dispersion of the spinon bands for various $\theta$ values. 

$\mathcal H_{spinon}$ as in eq \eqref{hspinon} is invariant under a projective translation and six-fold rotation symmetries, while breaking time-reversal and mirror symmetry except at the special point $\theta=\pm \pi/6$. The projective symmetry group for the spinons reads:
\begin{align}
&T_x: \psi_r\rightarrow (-1)^{r_y}\psi_{r+\hat x},\nonumber\\
&T_y:\psi_r\rightarrow \psi_{r+\hat y},\nonumber\\
&C_6: \psi_r\rightarrow \begin{cases} \tau^x \psi_{C_6 r} & (C_6 r)_y\mod 2=0 \\ -(-1)^{(C_6 r)_x}i\tau^x \psi_{C_6 r}& (C_6 r)_y\mod 2=1 \end{cases},
\end{align}
where $C_6 r$ is the $C_6$ rotated position of the site at $r$.

The low-energy physics of the  CSL1 is captured by the following action:

\begin{equation}
    L=-\frac{1}{4\pi} \alpha_1 d \alpha_1 -\frac{1}{4\pi}\alpha_2 d \alpha_2+\frac{1}{2\pi} ad(\alpha_1+\alpha_2)+\frac{1}{2} \frac{A_s}{2} d (\alpha_1-\alpha_2)
\end{equation}
where $A_s$ is the external spin field in the convention that its charge $q_s=\frac{1}{2}$ for $S_z=\frac{1}{2}$. $\alpha_1,\alpha_2$ are introduced to represent the IQH of $f_{\uparrow}$ and $f_{\downarrow}$ respectively.  $a$ is the internal $U(1)$ gauge field shared by the slave boson $b$ and the spinon $f_\sigma$.

We can also integrate $a$ to lock $\alpha_1=-\alpha_2=\alpha$, then we get

\begin{equation}
\label{csl1action}
    L=-\frac{2}{4\pi}\alpha d \alpha+\frac{1}{2\pi}A_s d \alpha
\end{equation}

This is a $U(1)_2$ theory which also describes the $\nu=\frac{1}{2}$ Laughlin state. Exploiting the relation between spin $1/2$ and hard core bosons, one can map the $\nu=\frac{1}{2}$ Laughlin state of boson to a spin state, which corresponds to CSL1.  Thus the CSL1 has the same topological order as the $\nu=\frac{1}{2}$ Laughlin state. Indeed a model wavefunction using the Laughlin state was proposed by Kalmeyer-Laughlin\cite{kalmeyer1987equivalence}.  But we need to emphasize that a general state with any $\theta \neq \pm\frac{\pi}{6}$ is in the same topological phase as the Kalmeyer-Laughlin state. There are two anyons in this phase: $I$ and $s$, with $s$ as a spin $1/2$ semion. The edge state is characterized by chiral central charge $c=1$.

\subsection{CSL2: Projected $d+id$ superconductor}

The projected $d+id$ superconductor is given by the $u$ matrices below:
\begin{align}
\label{d+id}
&u_{r,r+\hat x}= \frac{1}{\sqrt{\chi^2+\eta^2}}(\chi \tau^z+\eta \tau^x),\nonumber\\
&u_{r,r+\hat y}=\frac{1}{\sqrt{\chi^2+\eta^2}}[\chi \tau^z+\eta( \tau^x \cos \frac{4\pi}{3}-\tau^y\sin\frac{4\pi}{3})],\nonumber\\
&u_{r,r+\hat x+\hat y}=\frac{1}{\sqrt{\chi^2+\eta^2}}[\chi \tau^z+\eta( \tau^x \cos \frac{2\pi}{3}-\tau^y\sin\frac{2\pi}{3})],
\end{align}
with $\chi,\eta$ as real parameters and explicit translation invariance. The pairings for $3$ bonds generated by $C_3$ rotation have phases $0,2\pi/3,4\pi/3$ respectively, the same as that of a $d+id$ superconductor

The topological order depends on the angular momentum of the corresponding pairing, as classified by Kitaev's sixteenfold way\cite{kitaev_2006}.  With $SU(2)$ spin rotation symmetry, the pairing can only be in the even angular momentum channel, making the  gauged $d+id$ superconductor as  the simplest chiral spin liquid in the CSL2 category.  
There are four anyons labeled by $1,e,m,\epsilon$ in CSL2\cite{moroz_2017}. $e,m$ are semions with $\pi/2$ self-statistics and trivial mutual statistics. $\epsilon$ is a bound state of $e,m$ and has fermionic statistics. This state corresponds to $\nu=4$ of Kitaev's sixteenfold classification of $Z_2$ topological order\cite{kitaev_2006} and its topological order is $U(1)_2\times U(1)_2$. Its edge theory has chiral central charge $c=2$.  The wavefunction of CSL2 is a Gutzwiller projection of a $d+id$ superconductor instead of a s wave superconductor.

\subsection{Relation between two CSLs}

We note that these two CSL states generically are not equivalent, since the invariant gauge group (IGG) for projected $d+id$ states(CSL2) is $Z_2$, while IGG for CSL1 is $U(1)$. Moreover, they differ in the topological orders and anyon contents as described above.  

We note however the following collision between these two ansatz and clarify its meaning. For CSL2, at the special point when $\eta=\sqrt{2} \chi$, the Wilson loop around one triangle for projected $d+id$ reads $\Phi=u_{r,r+\hat x}u_{r,r+\hat y}u_{r,r-\hat x-\hat y}\propto i \tau^0$, equivalent to that of the CSL1 at $\theta=0$. Hence the two states are gauge equivalent at the special point.
Specifically, there is an SU(2)  gauge transform $g_r$ which rotates the mean field ansatz for CSL1 \eqref{hspinon} at $\theta=0$ to the mean field $d+id$ ansatz of CSL2 eq \eqref{d+id} at $\eta=\sqrt{2}\chi$:

\begin{align}
\label{csltransform}
&\psi_{r}\rightarrow g_{r} \psi_{r},\nonumber\\
& g_{r}=\begin{cases} 1& r_x \textrm{ even, }r_y \textrm{ even}\\
-\frac{\tau^z+\sqrt{2} \tau^x}{\sqrt{3}} & r_x \textrm{ odd, }r_y \textrm{ even}\\
\frac{-(\tau^z+\sqrt{2} ( \tau^x \cos \frac{4\pi}{3}-\tau^y\sin\frac{4\pi}{3}))}{\sqrt{3}}& r_x \textrm{ even, }r_y \textrm{ odd}\\
 \frac{-i(\tau^z+\sqrt{2} ( \tau^x \cos \frac{2\pi}{3}-\tau^y\sin\frac{2\pi}{3}))}{\sqrt{3}} &r_x \textrm{ odd, }r_y \textrm{ odd.}\end{cases}
\end{align}

At the special point $\theta=0$, the IGG is actually enlarged to  $SU(2)$ and the low energy theory is $SU(2)_1$, which is known to be equivalent to the $U(1)_2$ theory.  Hence the special point with $\theta=0$ also belongs to the CSL1. At the same time, the point $\eta = \sqrt 2 \chi$ does not belong to CSL2, since the gauge group is enlarged to SU(2). However, on moving away from this special point, a Higgs condensate develops, lowering the gauge group to Z$_2$.

One can ask what the gauge transformation above accomplishes when $\theta$ shifts away from $0$ for CSL1 ansatz. Then the above rotation eq \eqref{csltransform} on the CSL1 does not restore explicitly lattice translation as present in the $d+id$ ansatz (see appendix \ref{app:csltrans}). The transformed state at $\theta\neq 0$ has $d+id$ pairing, but breaks translation symmetry and despite the fact that it appears to be a pairing state, it secretly possesses a U(1) IGG.

\begin{figure}
 \captionsetup{justification=raggedright}
    \centering
     \adjustbox{trim={.13\width} {.5\height} {.35\width} {.\height},clip}
   { \includegraphics[width=0.8\textwidth]{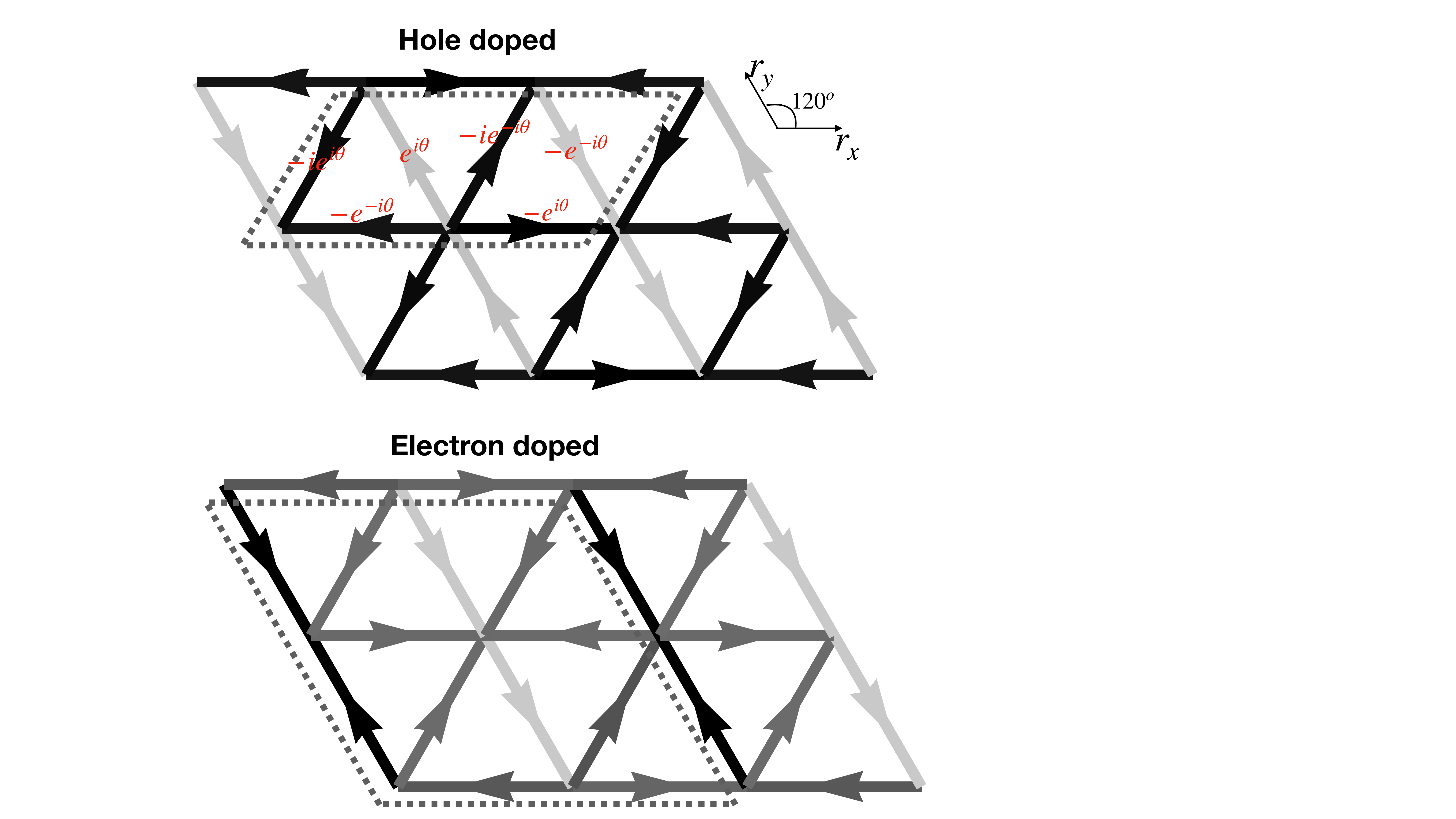}}
 \caption{A doubled unit cell shown in dotted parallelogram. Spinon hopping ansatz of the CSL shown in red on the bond inside the dotted parallelogram, direction in line with arrows on bonds.  The relative current strength $I_{bond\langle ij\rangle}=Im[c^\dagger_ic_j]$ of the chiral metal upon infinitesimal doping of holes ($\theta=0$) shown schematically by the bond gray level (the smaller the current is, the grayer the bond looks) with arrows indicating current directions. The currents are given by holon condensation at $\Gamma,M_2$ momenta such that the kinetic energy of electrons $- t\sum_{\langle ij\rangle} c_i^\dagger c_j$ is optimized. For doping holes, holons fully condense at $M_2$ and the currents have a \emph{doubled} unit cell structure. For doping electrons, currents are opposite to those in the hole scenario.}
    \label{fig:chiral_metal}
\end{figure}

\section{Chiral metal}
\label{chiral_metal}
At hole doping level $x$, the average density of the holon and the spinon per site becomes  $n_i^b=x$ and $n_i^f=1-x$.  One simple possibility is that slave boson condenses at zero temperature. This is obtained self-consistently for a mean field ansatz of holons. Assuming an electron hopping term in the parent Hubbard model of the form $H_{KE} = -t \sum_{\langle ij\rangle} c^\dagger_{i\sigma} c_{j\sigma} +{\rm h.c.}$ with real $t>0$, we can obtain the mean field theory of the holon as 
\begin{align}
    \mathcal H_{holon}=t\sum_{\langle ij\rangle,\sigma} \langle f_{i,\sigma} f_{j,\sigma}^\dagger\rangle b_i^\dagger b_j,
\end{align}
where $\langle f_{i,\sigma} f_{j,\sigma}^\dagger\rangle$ is computed on the ground state of the spinon mean field $\mathcal H_{spinon}$ and is proportional to spinon hopping terms on corresponding bonds.
Lowest energy states for $\mathcal H_{holon}$ lie at $\Gamma ,M_2$ momenta, and hence holon condensation is generically a superposition of such two states. The favored condensation is given by optimizing the kinetic energy of electrons in $H_{KE}$,
\begin{equation}
E_{KE}=t\sum_{\langle ij\rangle,\sigma} \langle f_{i,\sigma}f_{j,\sigma}^\dagger\rangle \langle b_{i}^\dagger\rangle \langle b_j\rangle,
\end{equation}
where the holon condensed values
\begin{eqnarray}
\label{holonwfn}
\langle b_{i} \rangle=\sqrt{2x}[p_\Gamma u_\Gamma(i_s)+(1-p_\Gamma)(-1)^{i_y}u_{M2}(i_s)],
\end{eqnarray}
where $p_\Gamma\in [0,1]$ is the fraction of holons that condense at $\Gamma$. $u_{\Gamma,M2}(i_s=A,B)$ are the Bloch wavefunctions at $\Gamma,M_2$ momenta on two sublattices,
 \begin{eqnarray}
u_\Gamma(A)=0.63(1+i),u_\Gamma(B)=0.46,\nonumber\\
u_{M2}(A)=0.33(1-i),u_{M2}(B)=0.89,
\end{eqnarray} 
identically for any $\theta$.

We found that for negative hopping in $H_{KE}$, the holons fully condense at $M_2=(0,\pi)$ to optimize the kinetic energy, i.e. $p_\Gamma=0$ in eq \eqref{holonwfn}. This is true for any $\theta\in (-\pi/6,\pi/6)$. For positive hopping (or equivalently doping electrons), holons fully condense at $\Gamma$,i.e. $p_\Gamma=1$. The wave functions with minimal energy for $-\mathcal H_{holon}$ read,
\begin{eqnarray}
   \tilde u_\Gamma(A)=0.33(-1-i),\tilde u_\Gamma(B)=0.89,\nonumber\\
  \tilde u_{M2}(A)=0.63(-1+i),\tilde u_{M2}(B)=0.46.
\end{eqnarray}

Holon condensation leads to the electron operator as 
 \begin{equation}
 \label{ef}
 c_{i,\sigma}=\langle b_i^\dagger \rangle f_{i,\sigma}.
 \end{equation}
Spinons are away from integer filling upon doping and states at the valence band top are emptied as Fermi pockets. Electrons, identified as spinons in eq \eqref{ef}, hence partially fill states in a band and form a metal with $\pi$ flux and breaks translation, rotation and time reversal, as the spinon Hamiltonian does.  Numerically, we found a nonvanishing bond current pattern of the metal $I_{bond\langle ij\rangle}\equiv Im[c_i^\dagger c_j]$ shown in figure \ref{fig:chiral_metal}(parameter $\theta=0$), at infinitesimal doping. 

For the electron doping scenario, one changes the parton definition to $c^\dagger_{i,\sigma}=b_i^\dagger f_{i,\sigma}$, which corresponds to a particle-hole transformation for the electrons, and hence results in a minus sign multiplying the hopping amplitude $-t$ in $H_{KE}$. To optimize $-H_{KE}$, holons all condense at $\Gamma$, currents are opposite to the hole doping scenario shown in fig \ref{fig:chiral_metal}, up to a $\hat r_x$ translation of $y$ bond currents.  We also present the current patterns at $\theta=\pm 0.2,\pm 0.4$ upon infinitesimal hole doping in supplementary fig \ref{fig:chiral_metal_cur_sup} with holons condensing fully at $\Gamma$.\footnote{In numerics, the  relative strength of the current may change by as much as 0.08 (normalized by the strongest current), when increasing the density of momenta used for calculations, from a density of 400 points per unit area to 10000 points per unit area.}The current pattern generically holds a $2\times 2$ unit cell due to holons condensing at both $\Gamma,M_2$ when going beyond a mean-field treatment; only when holons condense solely at $\Gamma$ or $M_2$ does it yield a $2\times 1$ unit cell pattern. 

The electrons have a non-vanishing Hall response as the spinons are in a Chern band. From Ioffe-Larkin rule, one has for the resistivity tensors $\rho_c=\rho_b+\rho_f$ where $\rho_b=0$ identically since holons have condensed. At small doping, $\rho_c=\rho_f$. Note that when holons condense, an extra term appears for the spinon mean field after substituting electron operators from eq \eqref{ef},
\begin{equation}
\overline{\mathcal  H_{spinon}}=\mathcal H_{spinon}-\sum_{\langle ij\rangle,\sigma}t \langle b_i\rangle \langle b_j\rangle^* f_{i,\sigma}^\dagger f_{j,\sigma},
\end{equation}
which should be small when light doping and we ignored such terms.

\begin{figure}
 \captionsetup{justification=raggedright}
    \centering
         \adjustbox{trim={.25\width} {.28\height} {.07\width} {.16\height},clip}
   { \includegraphics[width=0.65\textwidth]{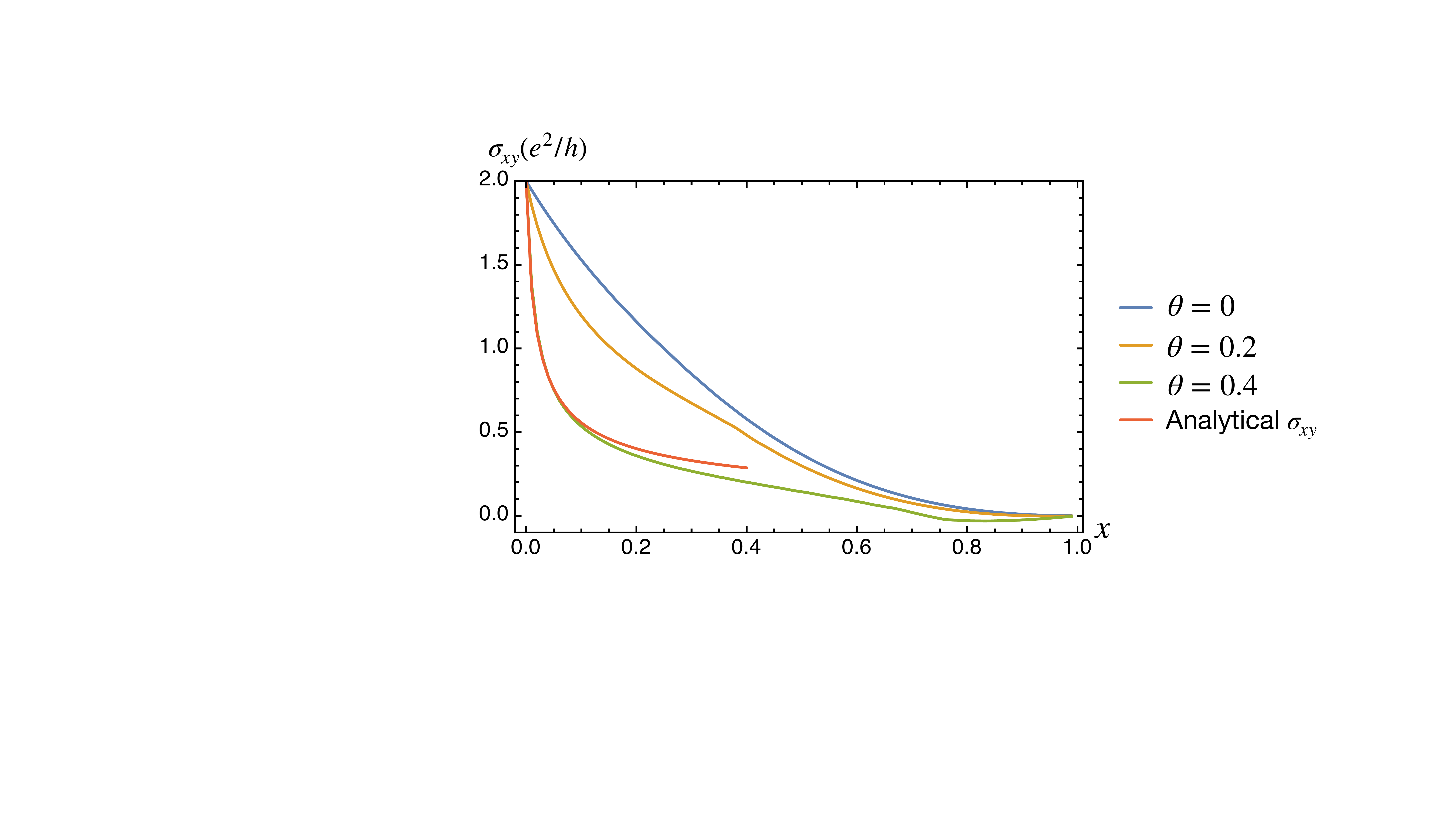}}
    \caption{The numerical hall conductivity $\sigma_{xy}$ of the chiral metal upon doping holes at fractions $x$ at selected values $\theta=0,0.2,0.4$, respectively. The results are expected to be accurate for light doping. The red line is the analytical expression for $\sigma_{xy}$ (eq \eqref{hall}) for $\theta=0.4$, near the gapless point $\theta=\pi/6$. It fits well with the numerical results at small doping $x<0.1$. 
    $\sigma_{xy}$ in the electron doping scenario is identical to that of the hole-doping case at same doping level $x$.
    }
    \label{fig:hall_metal}
\end{figure}

 For a free fermion system $\sigma_{xy}$ is given by the Kubo formula as
\begin{equation}
\sigma_{xy}=\frac{e^2}{\hbar}\int_{E_k<\epsilon_f} \frac{d^2k}{4\pi^2} 2 Im[\langle \frac{\partial u}{\partial k_1}|\frac{\partial u}{\partial k_2}\rangle],
\end{equation}
where $u(k_1,k_2)$ is the periodic Bloch wavefunction for $\mathcal H_{spinon}$ and the integration is over filled states. See Appendix \ref{app:hall} for numerical details on calculating Berry curvatures.  Fig. \ref{fig:hall_metal} shows the hall conductivity $\sigma_{xy}$ as one varies hole doping fraction $x$ and hopping phase $\theta$. It extrapolates to the quantized value of $2e^2/h$ at zero doping. The hall conductivity upon doping electrons, meanwhile, is identical to that on the hole side of the same doping level $x$.

Near the gapless point $\theta=\pm\pi/6$, one could approximate states near valence band top as Dirac fermions with a chiral mass, and analytically (see Appendix \ref{app:hall}) express the hall conductivity by integrating Berry curvatures of filled states, i.e. the hall conductivity as a function of $\Delta=|\pi/6-\theta|,x$ reads
\begin{equation}
\label{hall}
\sigma_{xy}(x,\Delta)=\frac{4\sqrt{3}\Delta}{\sqrt{12 \Delta^2+4\sqrt{3}\pi x}}.
\end{equation}

\begin{figure}
 \captionsetup{justification=raggedright}
    \centering
         \adjustbox{trim={.0\width} {.0\height} {.0\width} {.0\height},clip}
   { \includegraphics[width=0.5\textwidth]{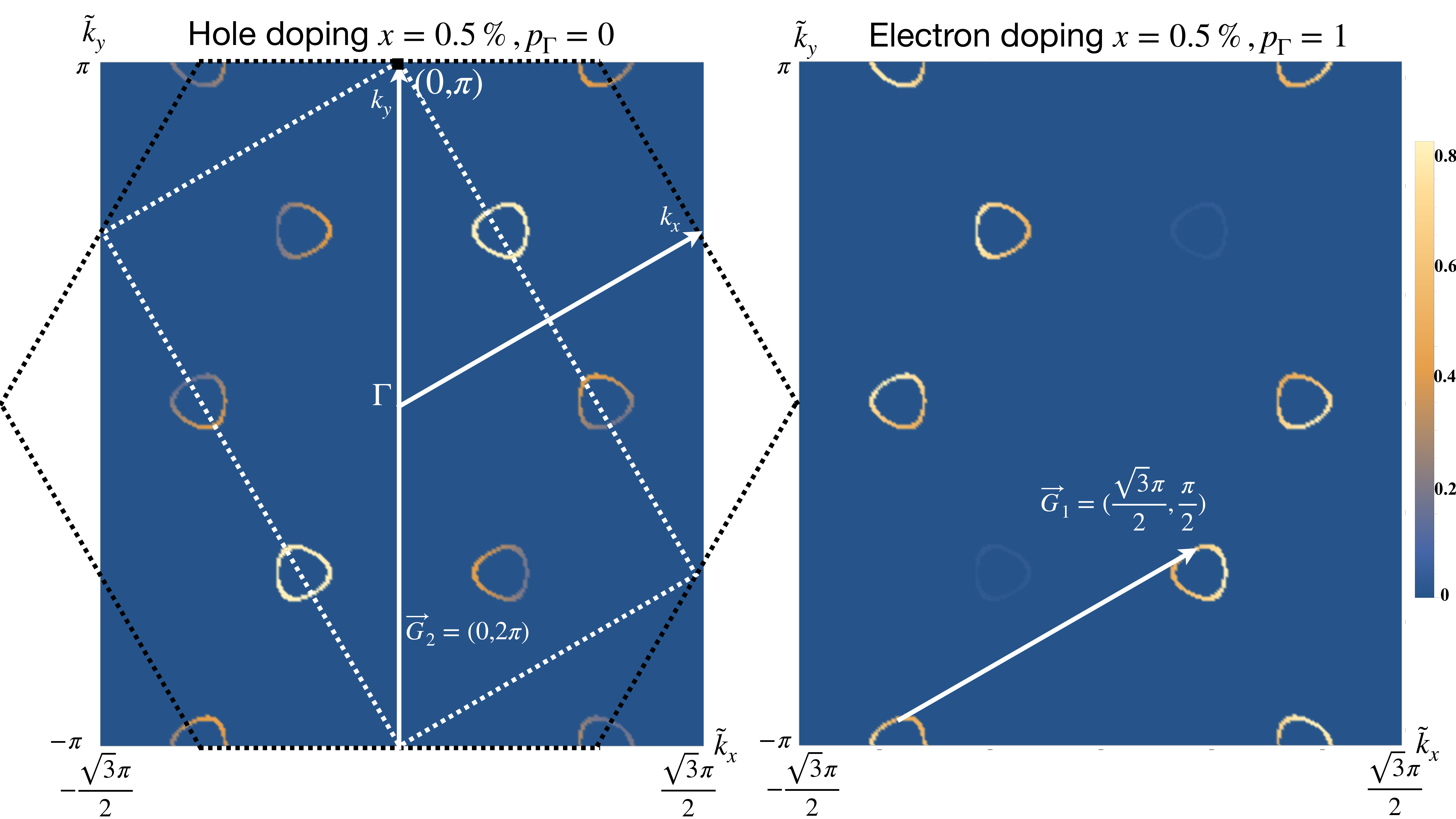}}
    \caption{The numerical results for spectral function for the chiral metal at zero frequency $A(\vec k,\omega=0)=2Im[G_R(\vec k,\omega=i0^+)]$(a.u.), where the retarded Green's function $G_R(\vec k,\omega)=\int_0^\infty \frac{dt}{2\pi}e^{i\omega t}\langle[c^\dagger_k(t),c_k(0)]\rangle_{GS}$, at $\theta=0$ and doping level marked at the top. The left(right) panel shows the case for hole(electron) doping with $p_\Gamma,1-p_\Gamma$ of holons condensing at $\Gamma,M_2$, respectively. The spectral function could be measured as in angle-resolved photo emission spectroscopy (ARPES), where the absolute momenta in the original  Brillouin zone(black dotted hexagonal) are resolved. The Fermi surface  is  duplicated after a translation of $\vec{G_1}$(coordinates written in $\tilde k_x-\tilde k_y$ basis) corresponding to the reciprocal vector of the reduced BZ(white dotted square in the left panel).}
    \label{fig:spectral}
\end{figure}

The enlarged unit cell implies a density wave formation in the chiral metal where states at $\vec k$ and $\vec k+\vec G_1$ couple together, where $\vec G_1=(\pi,0)$ is a reciprocal vector for the reduced Brillouin zone ($2\vec G_1$ for the original hexagonal Brillouin zone). Fig \ref{fig:spectral} presents numerical results on the spectral function $A(\vec k,\omega=0)=2Im[G_R(\vec k,\omega=i0^+)]$ where $G_R$ is the retarded Green's function, as measured by angle-resolved photo emission spectroscopy (ARPES). Details of calculation are shown in appendix \ref{app:spectral}. The duplication of Fermi surfaces separated by $\vec G_1$ is a signature of the density wave.

In the previous discussion we determine the ansatz of the spinon $f$ and the wavefunction of the holon $\langle b_i \rangle=\mu_i$ through mean field calculation. In practice, one can also apply the variational monte carlo technique to determine these parameters by studying the following model wavefunction:

\begin{equation}
    \Psi(x_1,...,x_N)=P_G (\prod_{i} \mu_{i}^{1-n_i}) \text{Slater}[x_1,...,x_N]
\end{equation}
where $\text{Slater}$ is a Slater Determinant according to the mean field ansatz of the spinon $f$. $P_G$ is the usual Gutzwiller projection to forbid the double occupancy.  Here we  introduce $\langle b_i \rangle=\mu_i$ as variational parameters. In practice one can focus only on a $2\times 2$ unit cell and hence there are four complex parameters for $\mu_i$.

The CSL1 with $\theta=0$ needs special treatment. At this special point, the spinon ansatz is gauge equivalent to a $d+id$ superconductor. Thus a translation invariant $d+id$ superconductor is also possible upon doping. However, there is no reason to expect that a CSL1 state is fine tuned to be at $\theta=0$.  For example, CSL1 found in $J_1-J_2-J_\chi$ model\cite{Gong_2017} is believed to be close to a Dirac spin liquid at $J_\chi=0$ line. Thus the spinon ansatz should be a Dirac spin liquid with small chiral mass, for which $\theta$ should be close to $\frac{\pi}{6}$. For generic $\theta$, a translation invariant $d+id$ superconductor is not possible in the picture of holon condensation.

A comparison with doping square lattice CSL is in order: On the square lattice, there are also two different CSL: CSL1 and CSL2 as on triangular lattice.  But for square lattice, the ansatz of spinons for   CSL1  can always be written in the form of $d+id$ superconductor (without hopping on next nearest neighbor) (please see Appendix.~\ref{app:square}).  In this case, doping can lead to a translation invariant $d+id$ or a chiral metal, depending on which one is energetically favorable.  This is different from triangular lattice where only a special point of CSL1 has spinons in a $d+id$ superconductor ansatz.

\section{Topological Superconductor from doping the generic Kalmeyer-Laughlin Spin Liquid CSL1}
\label{secnovelsc}
From the $Z_2$ spin liquid, RVB theory predicts a superconducting phase upon doping. One natural question is: can we also get a superconductor from doping a chiral spin liquid.  For CSL2, this is obvious as the spinons are already in a BCS state.  For the CSL1, the  focus here, it is not obvious how to do this except at the special point $\theta=0$.  At a generic point with $\theta \neq 0$, there is no gauge transformation which can make the spinon ansatz into a translation invariant $d+id$ superconductor.  There is a particular gauge in which  spinon are in a translation symmetry {\em breaking} superconductor, as shown in Appendix.~\ref{app:csltrans}.  However, in numerics the $d+id$ superconductor that emerges from doping a  CSL1 is found to be translation invariant\cite{jiang_2020}. This result therefore cannot be explained by the conventional RVB picture and requires a new theory.  In the following we propose precisely such an exotic mechanism for obtaining a topological superconductor from doping the Kalmeyer-Laughlin  CSL1.

We first turn to the slave boson theory $c_{i;\sigma}=b_i f_{i;\sigma}$ with the constraint $n_i^b=n_i^f$.  The slave boson $b$ and the spinon $f$ now have opposite gauge charges.   At zero doping, there is an internal flux $\Phi=\pi$ per unit cell and the effective filling in terms of the magnetic flux for $b$ and $f$ are $\nu_b=-2$ and $\nu_f=2$ respectively.   For the CSL, $f$ is in the $C=2$ Chern insulator, while the slave boson $b$ is in a trivial Mott insulator with one particle per site.   When we decrease the Hubbard $U$ or change the doping, the slave boson $b$ can not be in a Mott insulator anymore and will delocalize after a Mott transition.  The usual story is that the slave boson will just condense, which in our case leads to a Chern insulator at zero doping and chiral metal at finite doping.  However, as the slave boson is at filling $\nu_b=-2$ in terms of the magnetic flux, there is another option after the Mott transition:  the slave boson can be in a quantum Hall liquid phase instead of a superfluid phase.  The most natural choice is the bosonic integer quantum Hall (bIQH) state, which has been found numerically in lattice model with  flux $\Phi=\pi$ per unit cell\cite{yin_chen_2015}.   As we show later, the resulting phase turns out to be a topological superconductor in the same class of $d+id$ superconductor.   The easiest way to see  the superconductivity is through the Ioffe-Larkin rule: the resistivity tensor of electron is $\rho_c=\rho_b+\rho_f$, where $\rho_b,\rho_f$ are the resistivity tensor for $b$ and $f$.   As $b$ and $f$ are in quantum Hall phase with opposite Hall conductivity, $\rho_c=\begin{pmatrix}0&-2 \\ 2 &0 \end{pmatrix}+\begin{pmatrix}0&2 \\ -2 &0 \end{pmatrix}=\begin{pmatrix}0&0 \\ 0 &0 \end{pmatrix}$, which descrbes a superconductor. 

A superconductor phase obviously should survive when changing the doping. To keep it in a superconductor phase, we need both $b$ and $f$ in the quantum Hall phase even after doping.  At finite doping $x$, it turns out that the internal magnetic flux needs to adjust itself to:

\begin{equation}
 \Phi(x)= \frac{\langle da \rangle}{2\pi}=\frac{1-x}{2}.
\end{equation}

 In the language of PSG this means that $b$ and $f$ are obeying a magnetic translation symmetry: $T_1 T_2=T_2 T_1 e^{\pm i 2\pi\Phi(x)}$, where $\pm$ arises because $b$ and $f$ carry opposite gauge charges.     The electron $c=bf$, however, still satisfies the usual translation symmetry because the phase factors from $b$ and $f$ cancel each other.  We believe that the final electron phase is translation invariant.

The lock of the internal flux to the additional doping keeps the spinon $f$ in the $C=2$ Chern insulator phase  and $b$ in the bIQH phase according to the streda formula at arbitrary doping level $x$. The final wavefunction is a product of two quantum Hall wavefunctions:

\begin{equation}
\Psi(\vec{x}_1, ...,\vec{x}_N)= P_{G} \Psi_{bIQH}(\vec{x}_1,...,\vec{x}_N) \Psi_{fIQH}(\vec{x}_1,...,\vec{x}_N)
\label{eq:model_wavefunction}
\end{equation}
where $\Psi_{bIQH}$ is the wavefunction of the bIQH state at $\nu=-2$ and $\Psi_{fIQH}$ is the wavefunction of a fermionic IQH state with $\nu=2$. $P_G$ is the usual Gutzwiller projection to forbid double occupancy.  A concrete wavefunction of $\Psi_{bIQH}$ based on a mean field ansatz can be found in Appendix.~\ref{appendix:model_wavefunction}.

One can see  that this model wavefunction is quite distinct from the usual RVB wavefunction.  In our wavefunction, none of $b$ and $f$ have pairing structure, but the resulting phase is a superconductor.   In the following we study the topological property of this phase more carefully.   The low energy theory is

\begin{equation}
    L=L_{b,IQH}+L_{f,IQH}
\end{equation}

We have

\begin{align}
   & L_{f,IQH}=-\frac{1}{4\pi}\alpha_1 d \alpha_1 -\frac{1}{4 \pi} \alpha_2 d \alpha_2\notag\\
   &+\frac{1}{2\pi} a d (\alpha_1+\alpha_2)+\frac{1}{2\pi}\frac{A_s}{2} d (\alpha_1 -\alpha_2)
    \label{eq:f_IQH}
\end{align}
and
\begin{equation}
    L_{b,IQH}=\frac{1}{4\pi}\beta_1 d \beta_2+\frac{1}{4\pi} \beta_2 d \beta_1 +\frac{1}{2\pi}(A_c-a) d (\beta_1+\beta_2)
    \label{eq:b_IQH}
\end{equation}
where we have used the fact that the $K$ matrix for bosonic IQH with $\sigma_{xy}=-2$ is $K=\left(\begin{array}{cc}0 & -1 \\ -1 &0 \end{array}\right)$ and  charge vector is $q=(1,1)^T$.  The IQH with $\sigma_{xy}=2$ for fermion is described by  $K=\left(\begin{array}{cc}1 & 0 \\ 0 &1 \end{array}\right)$ and  charge vector $q=(1,1)^T$.   $\frac{A_s}{2}$ is the spin gauge field. Here we use the convention that $q_s=\pm \frac{1}{2}$ for spin up and down. $A_c$ is the external EM field and we assign charge to the slave boson $b$. $a$ is the internal gauge field shared by the slave boson $b$ and spinon $f_{\sigma}$.

We can simplify the action by integrating out $a$ first, which locks $\beta_2=\alpha_1+\alpha_2-\beta_1$. After substitution, we get

\begin{align}
    L&=-\frac{2}{4\pi}\beta_1 d \beta_1 - \frac{1}{4\pi}\alpha_1 d \alpha_1 -\frac{1}{4\pi} \alpha_2 d \alpha_2 \notag\\
    &+\frac{1}{2\pi} A_c d(\alpha_1 + \alpha_2)+\frac{1}{2\pi} \frac{A_s}{2} d (\alpha_1 - \alpha_2)+\frac{1}{2\pi} \beta_1 d (\alpha_1+\alpha_2)
\end{align}

Relabeling $\alpha_c=\frac{\alpha_1+\alpha_2}{2}$, $\alpha_s=\frac{\alpha_1-\alpha_2}{2}$, $\beta=-\beta_1+\frac{\alpha_1+\alpha_2}{2}$, we can rewrite the action as

\begin{equation}
\label{dwavesc}
    L=-\frac{2}{4\pi}\beta d \beta -\frac{2}{4\pi}\alpha_s d \alpha_s +\frac{1}{2\pi} A_s d \alpha_s+\frac{2}{2\pi}A_c d \alpha_c
\end{equation}

We need to emphasize that the quantization of the charge is different from the usual Chern-Simons theory.  We have the following dictionary: $q_\beta=-q_{\beta_1}$, $q_c=q_1+q_2+q_{\beta_1}$ and $q_s=q_1-q_2$, where $q_{\beta_1}, q_1,q_2$ are the charges of $\beta_1,\alpha_1,\alpha_2$. We require $q_{\beta_1},q_1,q_2$ to be integers to satisfy the usual quantization rule of the Chern-Simons theory.  The excitation is labeled by $l=(q_\beta, q_c, q_s)$.

Given that there is no Chern-Simons term for $\alpha_c$, it represents a gapless charge mode and we have a superconductor.  $\alpha_c$ represents the Goldstone mode and its charge $q_c$ labels the vortex.  As the smallest $q_c=q_1+q_2+q_{\beta_1}$ is $\pm 1$, the elementary charge of the superconductor is  $Q=2\frac{d\alpha_c}{2\pi}=2$ and we conclude that it is a charge $2e$ superconductor.

The other excitations are listed in Table.~\ref{Table_list_of_anyons}. There are in total four different anyons $1, e, m, \epsilon$, where $e$ and $m$ are two semions and $\epsilon=e m$ is a fermion.    For each excitaion, we can define two charges. The first one is the spin $S_z=\frac{1}{2}q_s$. The second one is the vortex charge $V=q_c$. The statistics of the anyon is $\theta=\frac{\pi}{2}(q_s^2+q_\beta^2)$.    We find that the two semions $e,m$ always bind with vortex and cost infinite energy.  The only finite energy excitation is the fermion $\epsilon$, which can be identified as a Bogoliubov fermion.  $\epsilon$ has a  mutual statistics $\theta_{\epsilon e}=\theta_{\epsilon m}=\pi$ with $e,m$, consistent with the braiding of bogolibov fermion around an elementary vortex. The excitations and their topological properties match that of the $d+id$ superconductor.  There is also a spin quantum Hall conductivity $\frac{1}{2} \frac{1}{4\pi} A_s d A_s$, in agreement with the $d+id$ superconductor\cite{read_2000,senthil1999spin}.  In summary, we believe the  topological property and the symmetry quantum number of the state defined in Eq.~\ref{eq:model_wavefunction} is the same as the $d+id$ superconductor.

\begin{table}
\captionsetup{justification=raggedright}
\centering
\begin{tabular}{|c|c|c|c|c|c|}
\hline
&$(q_{\beta_1},q_1,q_2)$ & $(q_\beta, q_c, q_s)$ & $\theta$ & $S_z$ & $V$ \\ \hline
$I$& $(0,0,0)$ & $(0,0,0)$ & $0$  & $0$  & $0$ \\ \hline
 $e$& $(0,1,0)$ & $(0,1,1)$ & $\frac{\pi}{2}$  & $\frac{1}{2}$  & $1$ \\ \hline
 $m$& $(-1,0,0)$ & $(1,-1,0)$ & $\frac{\pi}{2}$  & $0$  & $-1$ \\ \hline
 $\epsilon$ & $(-1,1,0)$ & $(1,0,1)$ & $\pi$  & $\frac{1}{2}$  & $0$ \\ \hline
\end{tabular}
\caption{List of anyons in the topological superconductor.  $V$ is the vortex charge. $S_z$ is the spin quantum number and $\theta$ is the self statistics.  We have $V=q_c$, $S_z=\frac{1}{2} q_s$ and $\theta=\frac{\pi}{2}(q_\beta^2+q_s^2)$.}
\label{Table_list_of_anyons}
\end{table}

We comment on similarity and difference from our construction here and a previous theory on doped Dirac spin liquid on Kagome lattice\cite{ko_2009}. Both our proposal and that of Ref.~\onlinecite{ko_2009} needs internal magnetic flux adjusting to the doping.  The superconductor proposed in Ref.~\onlinecite{ko_2009} is not a BCS superconductor and it needs time reversal to be broken only after doping.  In our case, there is already an internal magnetic flux breaking time reversal in the Mott insulator and the proposed superconductor is in the same class as a $d+id$ superconductor from BCS theory. Hence our proposal may be easier to be realized in realistic models. Actually, a recent numerical study confirms the existence of a $d+id$ superconductor\cite{jiang_2020} from doping a  CSL1,  consistent with our theory.

\section{Deconfined superconductor-insulator transition}

\label{transition}
\begin{figure}
 \captionsetup{justification=raggedright}
    \centering
         \adjustbox{trim={.19\width} {.63\height} {.24\width} {.05\height},clip}
   { \includegraphics[width=0.8\textwidth]{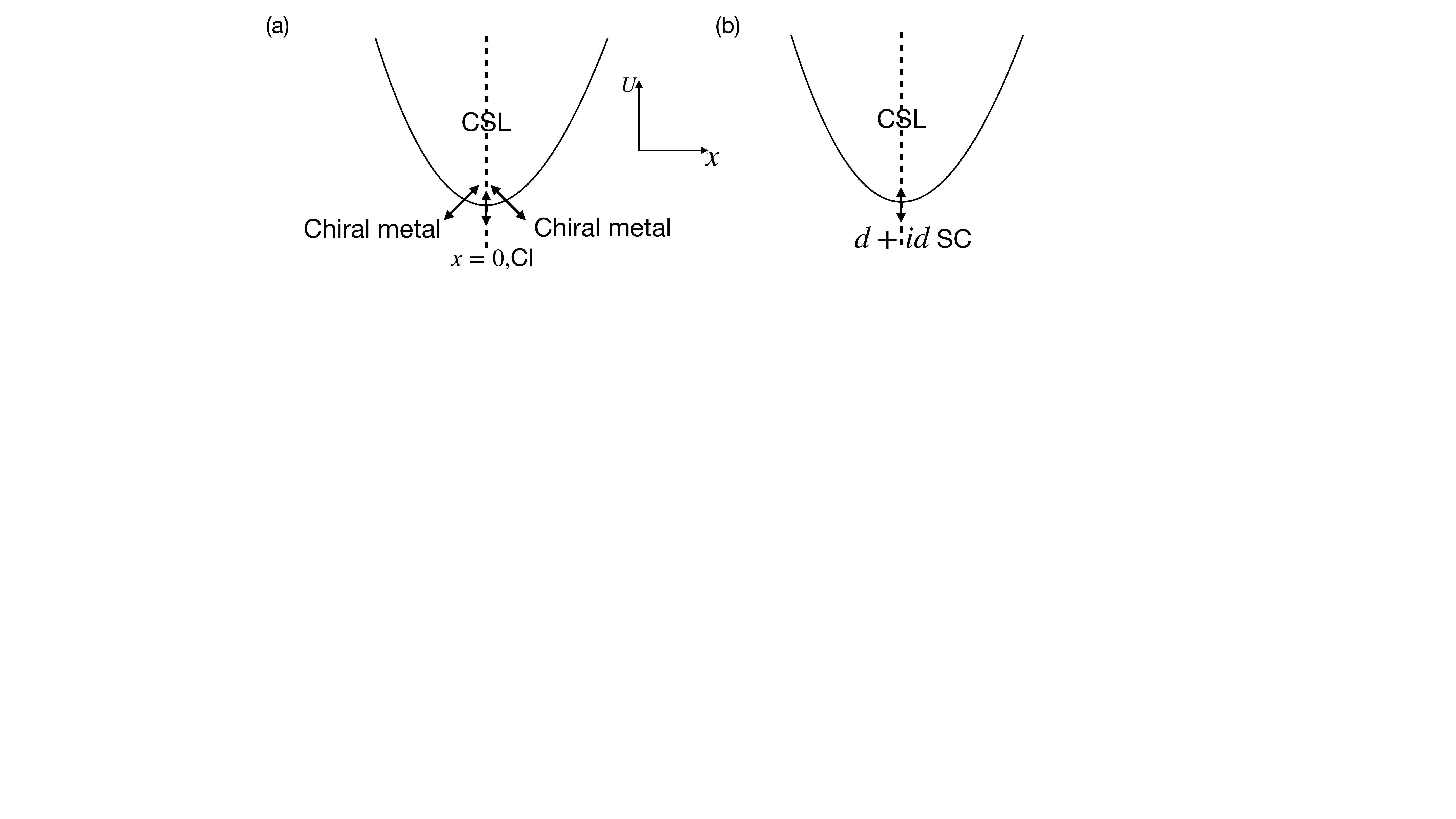}}
    \caption{The schematic phase diagram of the system as a function of interaction strength $U$ and doping level $x$ (positive,negative $x$ denotes doping electrons, holes, respectively). As one decreases $U$ or dopes the system, the CSL may become a chiral metal(left) or $d+id$ superconductor (SC)(right), depending on energetics. For the left figure, at exactly $x=0$ and small $U$, the system transits to a Chern insulator(CI) with Hall conductivity $\sigma_{xy}=2$.}
    \label{fig:uxphase}
\end{figure}

In this section we discuss the critical point between the chiral spin liquid Mott insulator and the chiral metal or the topological superconductor.  The transition can be tuned by either doping $x$ or via bandwidth control by changing the Hubbard $U$ as shown schematically in Fig \ref{fig:uxphase}. At finite doping, holons may remain in a Mott insulator, i.e., localized by disorder and hence the CSL phase extends to $x\neq 0$ in fig \ref{fig:uxphase}.  Usually the bandwidth controlled or doping controlled Mott transition is described by the condensation of the holons\cite{senthil_2008}.  The new feature in our theory is the possibility of an unconventional route: the transition into a bIQH insulator for the holons actually closes the Mott gap and makes the electrons into a topological superconductor phase.

\begin{figure}
 \captionsetup{justification=raggedright}
    \centering
     \adjustbox{trim={.2\width} {.35\height} {.32\width} {.3\height},clip}
   { \includegraphics[width=0.9\textwidth]{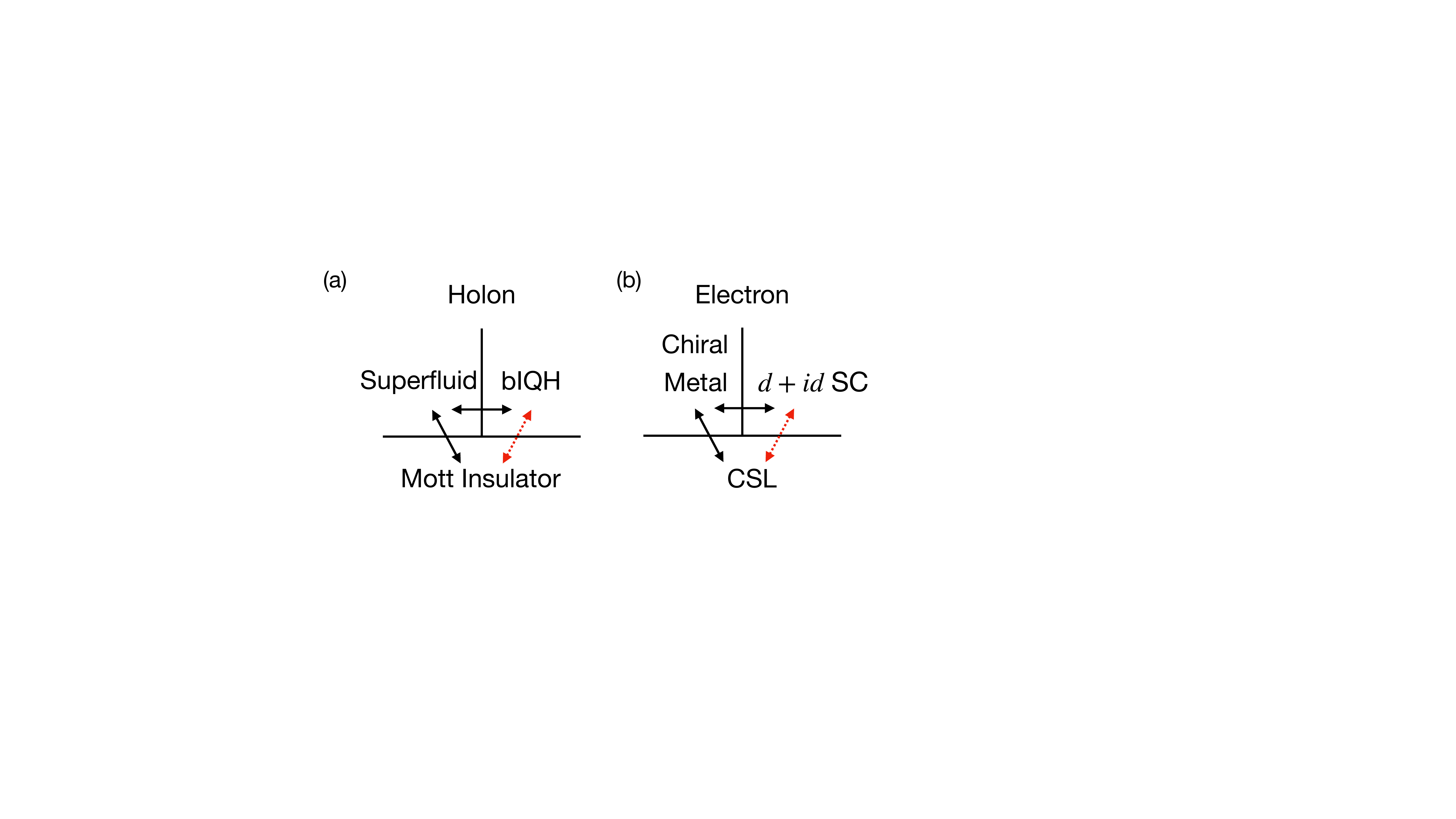}}
 \caption{The possible phases of holons and the corresponding physical phases. The transition of holons from Mott insulators or boson integer quantum hall (bIQH) to superfluid is described by Bose condenstion in $(2+1)D$. The direct transition from Mott insulators to bIQH marked by dotted arrows is studied in sec \ref{transition}.}
    \label{fig:comp_phase}
\end{figure}

The different electron phases associate with different  phases of holons, as shown in fig \ref{fig:comp_phase}. Therefore we can reduce the critical theory to the transition between different bosonic phases for the slave boson, similar to the theory in Ref.~\onlinecite{senthil_2008}. The transitions marked by black arrows in the fig \ref{fig:comp_phase}(a) from a trivial gapped state or an IQH state to a superfluid, are described by bose condensation.   The final transition for the corresponding electron phases in Fig.~\ref{fig:comp_phase}(b) needs to further include the gauge field.  At zero doping, the chiral metal is really a Chern insulator.    In this section we focus on the superconductor insulator transition marked by the red arrow in Fig.~\ref{fig:comp_phase}.

A key component of this superconductor insulator transition is the plateau transition for the slave boson. This transition marked by dotted red arrows from a Mott insulator to bIQH has been studied in Ref. \onlinecite{grover_2012,yuan_ming_2014,Barkeshli_2014}.  Using a parton construction with $6$ fermionic parton fields (see Appendix \ref{app:transition}), this transition from Mott to bIQH state happens when one particular parton $\psi_1$ band's Chern number changes from $C=1$ to $C=-1$, while other partons remain in some Chern bands. Hence the transition can be described by $2$ Dirac cones $\chi_{1,2}$ in the $\psi_1$ bands, whose mass sign changes induce changes in Chern number.   We will present the critical theory and analyze its universal property in this section.

\subsection{Critical action and phase diagram}
 
 The critical action reads from Appendix \ref{app:transition}
\begin{align}
\label{cri_action}
     \mathcal L_{cri}=\sum_{i=1,2} \bar\chi_i [\eta_\mu (i\partial_\mu+A_{c,\mu}+\gamma_\mu-a_\mu)\chi_i+m_i\bar \chi_i\chi_i\nonumber\\
    +\frac{1}{4\pi}(\beta d\beta-2\gamma d\beta-\beta_2d\beta_2+2(A_c+\gamma-a)d\beta_2)\nonumber\\
    -\frac{1}{4\pi} \alpha_1 d \alpha_1 -\frac{1}{4\pi} \alpha_2 d \alpha_2 \nonumber\\+\frac{1}{2\pi} (\frac{1}{2}A_s +a) d \alpha_1+\frac{1}{2\pi} (-\frac{1}{2}A_s +a) d \alpha_2,
\end{align}
where $\eta_\mu$'s are Pauli matrices acting on the Dirac spinor indices and we ignored the Maxwell terms for $A_c+\gamma-a$ coupled to the Dirac fermions. At the critical point, the $\chi_i$'s are gapless ($m_i=0$). The masses $m_1=m_2$ are enforced by the projective translation symmetry of the fermions. Properly integrating out $\chi_i$ when the masses are nonzero and $\chi_i$ filling a band with chern number $\pm 1$, requires a dual field $\beta_1$ and gives $\frac{\mp 1}{4\pi}\beta_1d\beta_1+\frac{1}{2\pi}\beta_1d(A_c+\gamma-a)$, respectively. 

When $m_1=m_2>0$, $\psi_1$ fills a band with $C_1=1$ and the theory describes a $d+id$ superconductor. Integrating out fields like $a, \chi_i,\gamma$ gives an action (appendix \ref{app:transition})
\begin{align}
    \mathcal L_{d+id}=\frac{1}{4\pi}(-2\alpha_cd\alpha_c+4\beta_2d\alpha_c-2\beta_2d\beta_2+2A_cd\alpha_c)\nonumber\\
   -\frac{2}{4\pi} \alpha_s d \alpha_s +\frac{1}{2\pi} A_s d \alpha_s,
\end{align}
with $\alpha_{c,s}$ related to $\alpha_{1,2}$ as in section \ref{secnovelsc}. If one identifies $\tilde \beta=\beta_2-\alpha_c$, the action is identical to the $d+id$ superconductor action in eq \eqref{dwavesc}.

When $m_1=m_2<0$,$\psi_1$ fills a band with $C_1=-1$ and the theory describes a chiral spin liquid (CSL1). Integrating out $\chi_i$ and $\gamma,\beta_2$ gives an action as eq \eqref{csl1action}(appendix \ref{app:transition}),
describing the response of CSL1 with the charge mode gapped.

\subsection{Universal properties of the critical point}
To obtain the transport properties of the critical theory, one integrates out $\alpha_c$ ($\alpha_{c,s}$ as defined in sec \ref{secnovelsc}) in the critical action \eqref{cri_action}, writing $\Gamma=\gamma-a$, and the critical theory reads,
\begin{align}
    \mathcal L_{cri}=\sum_{i=1,2} \bar\chi_i[\eta_\mu(i\partial_{\mu}+A_{c,\mu}+\Gamma_\mu)]\chi_{i}\nonumber\\
    -\frac{1}{4\pi}\Gamma d \Gamma
   +\frac{2}{4\pi}A_cd\Gamma+\frac{1}{4\pi}A_cdA_c\nonumber\\
 -\frac{2}{4\pi} \alpha_s d \alpha_s +\frac{1}{2\pi} A_s d \alpha_s.
\end{align}
The physical currents $J=\delta L_{cri}/\delta A_c$ are
\begin{eqnarray}
\label{current}
2\pi J_x=2\pi\mathcal J_x-E_y-e_y\nonumber\\
2\pi J_y=2\pi\mathcal J_y+E_x+e_x,
\end{eqnarray}
where we denote $\epsilon^{\mu\nu\rho}\partial_{\nu}A_{c,\rho}=(B,-E_y,E_x)$. In addition, we have
$\epsilon^{\mu\nu\rho}\partial_{\nu}\Gamma_\rho=(b,-e_y,e_x)$ and $\mathcal J_\nu$ as the current for $\chi$. The equations of motion for $\Gamma$ are
\begin{eqnarray}
2\pi\mathcal J_x+e_y-E_y=0,\nonumber\\
2\pi\mathcal J_y-e_x+E_x=0.
\end{eqnarray}

Combining the above relations and the assumed universal conductivity for $\chi$ at the critical point (QED-3), $\mathcal J_\nu=\sigma_\chi (E_\nu+e_\nu)$, one gets the conductivity from eq \eqref{current} as
\begin{eqnarray}
\label{sigma}
2\pi\sigma_c=\left (\begin{array}{cc} \frac{4\tilde\sigma_\chi}{\tilde\sigma_\chi^2+1} & \frac{-2(1-\tilde\sigma_\chi^2)}{\tilde\sigma_\chi^2+1}\\ \frac{2(1-\tilde\sigma_\chi^2)}{\tilde\sigma_\chi^2+1}& \frac{4\tilde\sigma_\chi}{\tilde\sigma_\chi^2+1}  \end{array}\right ),
\end{eqnarray}
where $\tilde\sigma_\chi=2\pi \sigma_\chi$.


In summary, the deconfined superconductor-insulator critical point should have a non-zero universal conductivity shown in Eq.~\ref{sigma}.  Given that both the superconductor and the CSL insulator have zero Hall conductivity, a non-zero Hall effect only at the critical point is quite remarkable. There is also a spin quantum Hall effect $\frac{1}{8\pi}A_s d A_s$, which remains constant across the transition.

\section{Conclusions and Outlook}

In summary, we propose two different phases from doping a $U(1)_2$ chiral spin liquid on triangular lattice.  One possibility is a chiral metal from simple slave boson condensation picture.  Another more exotic possibility is a topological superconductor obtained by putting both the slave boson and the spinon in a quantum Hall phase.  We argue that these two scenarios are consistent with two recent numerical studies of doped CSL in two different parameter regimes.  Our theory may be relevant to TMD heterobilayer where a spin 1/2 Hubbard model can be simulated on a triangular moir\'e superlattice\cite{wu2018hubbard,tang_2019,Regan2019optical,wang2019magic}.   It has also been proposed that a generalized CSL with different topological order  is possible in a $SU(4)$ model on moir\'e bilayer\cite{zhang_2020}. In future it is interesting to extend our theory of doped CSL to the case with $SU(N)$ spin. 

From a quantum criticality perspective, we propose an unconventional superconductor to Mott insulator transition, as described by a plateau transition of the charged holons.  If we start from a $d+id$ superconductor, there are two different Mott transitions towards two different chiral spin liquid  insulators.  We also find a deconfined critical point between a translation symmetry breaking Chern insulator and a translation invariant topological superconductor.  We hope to study this kind of deconfined critical points in more detail in the future. 

\section{Acknowledgement}
We would like to thank Ehud Altman, Olexei Motrunich and Zheng Zhu for useful discussions. We acknowledge funding from a  Simons Investigator award (AV) and the Simons Collaboration on Ultra-Quantum
Matter, which is a grant from the Simons Foundation (651440, AV). This research is funded  in part by the Gordon and Betty Moore Foundation's EPiQS Initiative, Grant GBMF8683 to AV.

\appendix
 \begin{widetext}

 \section{$SU(2)$ and $U(1)$ slave boson formalism}
\label{app:parton}
We remark that a full treatment of the parton representation for electrons have an $SU(2)$ gauge symmetry, which is given by two species of spin-$0$ holons $h_i=(b_{1,i},b_{2,i})^T$ and spinons $\psi_i=(\psi_{i,\uparrow},\psi_{i,\downarrow}^\dagger)^T$ to write the electron operators as
\begin{align}
c_{i,\uparrow}=\frac{1}{\sqrt{2}} h_i^\dagger\psi_i=\frac{1}{\sqrt{2}}(b_{1,i}^\dagger f_{i,\uparrow}+b_{2,i}^\dagger f_{i,\downarrow}^\dagger),\nonumber\\
c_{i,\downarrow}=\frac{1}{\sqrt{2}} h_i^\dagger(i\tau^y \psi_i^*)=\frac{1}{\sqrt{2}}(b_{1,i}^\dagger f_{i,\downarrow}-b_{2,i}^\dagger f_{i,\uparrow}^\dagger),
\end{align}
where both $\psi_i$ and $\bar\psi_i=i\tau^y\psi_i^*$ transform as doublets for $SU(2)$ group: $\psi_i\rightarrow U\psi_i,(U\in SU(2))$ and $\bar\psi_i\rightarrow U\bar\psi_i,(U\in SU(2))$. The physical electrons stay invariant under such $SU(2)$ transform given the holon doublet transform accordingly as $h_i\rightarrow Uh_i$. The gauge-invariant condition for physical Hilbert space (excluding double occupancy) constrain each site to contain an $SU(2)$ singlet, i.e. 
\begin{equation}
h_i^\dagger{\mathbf{\tau}}h_i+\psi_i^\dagger {\mathbf {\tau}}\psi_i|\Psi_{phys}\rangle=0,
\end{equation}
which for the $\tau^z$ component reads
\begin{equation}
b_{1,i}^\dagger b_{1,i}-b_{2,i}^\dagger b_{2,i}+\sum_s f_{i,s}^\dagger f_{i,s}=1.
\end{equation}

In the main text parton construction, we effectively fix the gauge for the holon sector as $b_{i,2}=0$ identically and leave out one holon $b_{i,1}$ as $b$ in the main text. This is justified since given the $SU(2)$ gauge group, one could rotate a state to one with $b_{i,2}=0$ without changing the physical state. The price is a reduced gauge group $U(1)$ away from half-filling, and that upon transforming the mean field for spinons with $SU(2)$ rotations,  the state changes physically upon doping. For instance, in eq \eqref{csltransform} one transforms the spinon ansatz from a $U(1)_2$ CSL1 to $d+id$ (CSL2), equivalent at half-filling, and yet upon doping holes, the state becomes a chiral metal and $d+id$ superconductor after condensing $b_1$, respectively. Conversely, one could fix the gauge for spinons and then different boson condensation leads to different physical states. In the paper we adopt the first approach, allowing only $b_1$ to condense, and hence different spinon ansatz, equivalent at half-filling, give different physical states upon doping. Therefore at $\theta=0$ on triangular lattice CSLs, upon doping, either a $d+id$ or a chiral metal could emerge upon condensing different holon states. This is also true on square lattice (see appendix \ref{app:square}), where a $d+id$ ansatz with zero diagonal hopping is equivalent to a chiral spin liquid (DSL with a chiral mass) through an $SU(2)$ transform. Hence on square lattice doping a CSL gives either $d+id$ superconductor or a chiral metal.

\section{Competing translation breaking pairing state on doping CSL1}
\label{app:csltrans}

At noted in the main text, at $\theta=0$, the $U(1)_2$ CSL1 is equivalent to a projected $d+id$ spin liquid on triangular lattice. Hence upon doping the $U(1)_2$ CSL1 at $\theta=0$, the resulting states are a competition between $d+id$ and chiral metal in sec \ref{chiral_metal}. When $\theta$ shifts from zero, however, one could still perform the gauge transform eq \eqref{csltransform}  on $U(1)_2$ CSL1 towards a $d+id$ pairing ansatz, which could be the competing pairing state with chiral metal upon doping the CSL1. Next we analyze the ansatz for such pairing state.

The CSL1 ansatz eq \eqref{hspinon} becomes $\tilde u_{ij}= g_j^\dagger u_{i,j} g_i$ after the transform,
\begin{align}
&\tilde u_{r,r+\hat x}=\frac{1}{\sqrt{3}}\left (\begin{array}{cc} e^{i\theta} & \sqrt{2} e^{-i\theta}\\ \sqrt{2} e^{i\theta} & -e^{-i\theta}\end{array}\right ), (r_x \mod 2=0)\nonumber\\
&\tilde u_{r,r+\hat x}=\frac{1}{\sqrt{3}}\left (\begin{array}{cc} e^{i\theta} & \sqrt{2} e^{i\theta}\\ \sqrt{2} e^{-i\theta} & -e^{-i\theta}\end{array}\right ), (r_x \mod 2=1)\nonumber\\
&\tilde u_{r,r+\hat y}=\frac{1}{\sqrt{3}}\left (\begin{array}{cc} e^{i\theta} & \sqrt{2} e^{-i\theta-i\frac{2\pi}{3}}\\ \sqrt{2} e^{i\theta+i\frac{2\pi}{3}} & -e^{-i\theta}\end{array}\right ), (r_x \mod 2=0)\nonumber\\
&\tilde u_{r,r+\hat y}=\frac{1}{3}\left (\begin{array}{cc} -i(e^{i\theta}+2e^{-i\theta+i\frac{2\pi}{3}}) & \sqrt{2}i (e^{-i\theta+i\frac{2\pi}{3}}-e^{i\theta})\\  \sqrt{2}i (-e^{i\theta-i\frac{2\pi}{3}}+e^{-i\theta}) & -i (e^{-i\theta}-2e^{i\theta-i\frac{2\pi}{3}})\end{array}\right ), (r_x \mod 2=1)\nonumber\\
&\tilde u_{r,r+\hat y+\hat x}=\frac{1}{\sqrt{3}}\left (\begin{array}{cc} e^{-i\theta} & \sqrt{2} e^{i\theta+i\frac{2\pi}{3}}\\ \sqrt{2} e^{-i\theta-i\frac{2\pi}{3}} & -e^{i\theta}\end{array}\right ), (r_x \mod 2=0)\nonumber\\
&\tilde u_{r,r+\hat y+\hat x}=\frac{1}{3}\left (\begin{array}{cc} i(e^{-i\theta}+2e^{i\theta-i\frac{2\pi}{3}}) & \sqrt{2}i (-e^{i\theta-i\frac{2\pi}{3}}+e^{-i\theta})\\ \sqrt{2}i (e^{-i\theta+i\frac{2\pi}{3}}-e^{i\theta}) & i (e^{i\theta}+2e^{-i\theta+i\frac{2\pi}{3}})\end{array}\right ). (r_x \mod 2=1)
\end{align}
One sees that at $\theta=0$, the above mean-field ansatz is identical to $d+id$ states in eq \eqref{d+id} at $\eta=\sqrt{2} \chi$, while $\theta\neq 0$, the pairing has $d+id$ symmetry yet the state breaks lattice translation, distinct from the conventional translation invariant $d+id$ states.

\section{Chiral spin liquids on square lattice}
\label{app:square}

For square lattice, under an appropriate gauge choice,  CSL mean-field ansatz in the spinor basis $\psi$ reads,
\begin{eqnarray}
\label{squared}
&u_{r,r+\hat x}=\cos (\theta)\tau^z+\sin(\theta)\tau^y,\nonumber\\
&u_{r,r+\hat y}=\cos(\theta)\tau^z-\sin(\theta)\tau^y,\nonumber\\
&u_{r,r+\hat x+\hat y}=\chi\cos (\theta)\tau^z+\eta\sin(\theta)\tau^x,\nonumber\\
&u_{r,r-\hat x+\hat y}=\chi\cos (\theta)\tau^z-\eta\sin(\theta)\tau^x, (\chi,\eta\in R)
\end{eqnarray}
where the IGG is reduced to $Z_2$ when $\theta\neq n\pi/4, (n\in Z)$ and $\chi\neq 0$. This can be seen by calculating Wilson loop $\Phi$ around a square or triangular loop on the square lattice,
\begin{align}
\label{squareflux}
&\Phi_{\square}=u_{r,r+\hat x}u_{r,r+\hat y}u_{r,r-\hat x}u_{r,r-\hat y},\nonumber\\
&=\cos 4\theta-i\sin 4\theta \tau^x,\nonumber\\
&\Phi_{\triangle}=u_{r,r+\hat x}u_{r,r+\hat y}u_{r,r-\hat x-\hat y}
=\chi(\cos 2\theta\cos \theta \tau^z\nonumber\\&+\sin 2\theta\cos \theta \tau^y)+\eta(\cos 2\theta\sin\theta \tau^x-i \sin 2\theta\sin\theta).
\end{align}

We note that the $d+id$ ansatz contain two different chiral spin liquids: (I) When $\theta=n \frac{\pi}{4}$ or $\chi=0$ with arbitrary $\theta$, the IGG is $U(1)$ and we get a CSL1 described by the $U(1)_2$ theory;  (II) For a generic ansatz  with diagonal hopping, the IGG is $Z_2$ and we have a CSL2 phase in the $\nu=4$ of the Kitaev's 16 fold way.

We note that $d+id$ ansatz \emph{ contains} all square $U(1) $ DSL with a chiral mass states (CSL1). The ansatz for $U(1)$ CSL1 is a staggered flux one with diagonal hopping,
\begin{eqnarray}
&u_{r,r+\hat x}=\cos (\theta)\tau^z+i(-1)^{r_x+r_y}\sin(\theta)\tau^0,\nonumber\\
&u_{r,r+\hat y}=\cos(\theta)\tau^z-i(-1)^{r_x+r_y}\sin(\theta)\tau^0,\nonumber\\
&u_{r,r+\hat x+\hat y}=(-1)^{r_x+r_y}\eta\sin(\theta)\tau^z,\nonumber\\
&u_{r,r-\hat x+\hat y}=-(-1)^{r_x+r_y}\eta\sin(\theta)\tau^z, (\eta\in R)
\end{eqnarray}
where IGG is $U(1)$ generated by $\tau^z$. By a gauge transform, such ansatz is equivalent to $d+id$ with $\chi=0$ in eq \eqref{squared}. The gauge transform reads,
\begin{eqnarray}
\psi_r\rightarrow \frac{1}{\sqrt{2}}(1+(-1)^{r_x+r_y} i\tau^y)\psi_r.
\end{eqnarray}

Hence a generic CSL1 phase on square lattice can be written in the $d+id$ ansatz. After doping a CSL1, we can get a translation invariant $d+id$ superconductor from the simple holon condensation picture, in contrast to triangular lattice.

\section{Numerical procedure for Berry curvatures and Dirac fermions near $\theta=\pi/6$}
\label{app:hall}
We present some numerical details for Berry curvatures and analytical derivation of Dirac Hamiltonian near $\theta=\pi/6$ gapless point.

Berry connections for Bloch wavefunctions $u(\vec k)$ for the electrons can be approximated numerically as $A_\epsilon d \vec\epsilon=Arg \langle u(\vec k+\vec \epsilon)|u(\vec k)\rangle$ where $\vec\epsilon$ is a small vector increment in momentum space and Arg gives the phase of the complex number. We mesh the $kx-ky$ plane with increments $\vec\epsilon_1=(\delta,0)$,$\vec\epsilon_2=(0,\delta)$, where $k_x,k_y$ are measured by the reciprocal vectors for basis vector $\hat x,\hat y$ in the main text, shown in fig \ref{fig:kspace} (b) and $\delta$ at the order of $0.01$ radian. The total Berry curvature on one elementary plaquette of the momentum mesh is given by (shown in Fig \ref{fig:kspace} (a))
\begin{eqnarray}
&\mathcal F(\vec k)d^2 k=Arg \langle u(\vec k+\vec \epsilon_1)|u(\vec k)\rangle\nonumber\\
&+Arg \langle u(\vec k+\vec \epsilon_1+\vec\epsilon_2)|u(\vec k+\vec\epsilon_1)\rangle\nonumber\\&+Arg \langle u(\vec k+\vec \epsilon_2)|u(\vec k+\vec \epsilon_1+\vec\epsilon_2)\rangle\nonumber\\
&+Arg \langle u(\vec k)|u(\vec k+\vec\epsilon_2)\rangle.
\end{eqnarray}
Note that for bands with nonzero Chern number, Berry curvature on some plaquettes in k-space may have discontinuity of $\pm 2\pi$, since total Berry curvature of one band in the above discretized formula is always zero. One manually makes up for the $2\pi$ discontinuities and then sum the Berry curvature totals $\mathcal F(\vec k)d^2 k$ over filled $\vec k$ states for hall conductivity of the spinons.

\begin{figure}
 \captionsetup{justification=raggedright}
    \centering
         \adjustbox{trim={.19\width} {.4\height} {.4\width} {.31\height},clip}
   { \includegraphics[width=0.95\textwidth]{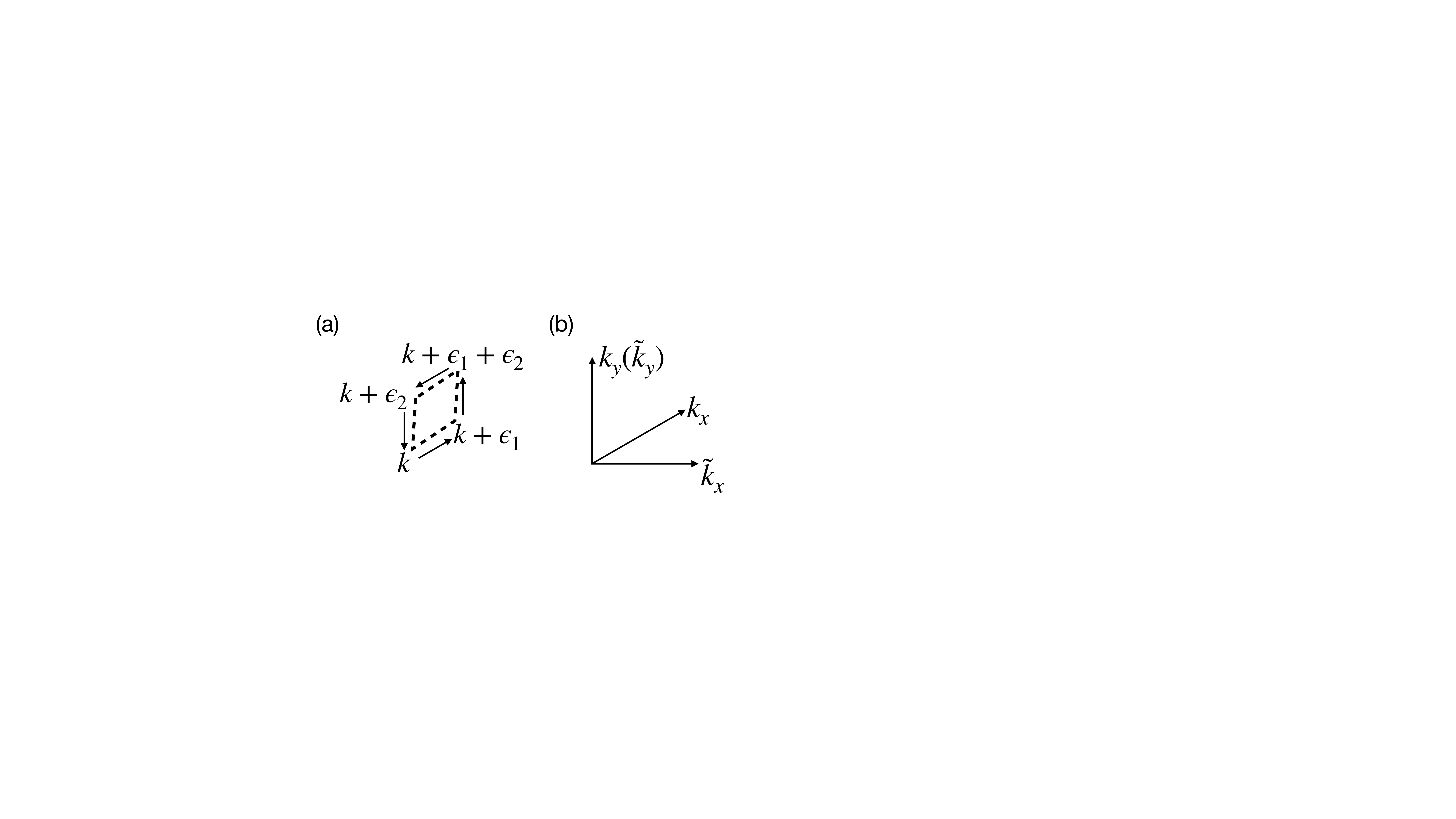}}
    \caption{(a)The numerical procedure to calculate Berry curvature of Bloch wavefunction in k-space. One circles around an elementary plaquette (dashed diamond) and adds up the phase of overlap of wavefunctions at the ends of each bond. (b) The reciprocal vectors for $\hat x,\hat y$ lattice vectors used in main text and the orthogonal coordinates used for Dirac hamiltonian eq \eqref{hdirac} $\tilde k_x,\tilde k_y$. }
    \label{fig:kspace}
\end{figure}

At $\theta$ close to $\pm \pi/6$, in the \emph{spinon} Hamiltonian, Dirac cones with a small mass emerge near half-filling.

We derive the Dirac hamiltonian near the gapless point. The spinon hamiltonian  eq \eqref {hspinon} in k-space reads
\begin{eqnarray}
\label{hspinonk}
\mathcal H_{spinon}=\sum_k \psi_k^\dagger ( -2 \cos (\theta+k_y) \tau^z\nonumber\\
-2\cos(\theta+k_x)\tau^x+2\cos(\theta-k_x-k_y)\tau^y)\psi_k,
\end{eqnarray}
 where $\psi_k=(f_{\uparrow,k},f_{\downarrow,-k}^\dagger)^T$. The k-space hamiltonian can be Taylor expanded near the gapless point, e.g. $\theta=\pi/6,(k_x,k_y)=(\pi/3,\pi/3)$ and we transform to orthogonal momentum coordinates shown in fig \ref{fig:kspace}(b) $\tilde k_x= \frac{\sqrt{3}}{2} k_x,\tilde k_y=k_y+k_x/2$, 
 \begin{eqnarray}
 \label{hdirac}
 \mathcal H_{Dirac}&= 2[(\frac{2}{\sqrt{3}} \tau^x-\frac{1}{\sqrt{3}}\tau^z-\frac{1}{\sqrt{3}} \tau^y)\tilde k_x+(\tau^z-\tau^y) \tilde k_y\nonumber\\
 &+(\theta-\frac{\pi}{6})(\tau^z+\tau^x+\tau^y)],
 \end{eqnarray}
 with the matrices multiplying $\tilde k_x,\tilde k_y,\sqrt{\frac{3}{2}}(\theta-\pi/6)$ obeying the anti-commutation relations of gamma matrices for Dirac hamiltonian in $(2+1)D$ up to an overall factor $2\sqrt{2}$. Hence indeed the band can be approximated by a Dirac fermion with a small mass. Another Dirac cone appears at $(\pi/3, -2\pi/3)$ in k-space with the same chirality. Note the velocity is uniform,  $v=2\sqrt{2}$ and the mass $m=2\sqrt{3} (\theta-\pi/6)$.
 The Berry curvature of the two-component spinor system has been well studied: representing the components of the Dirac hamiltonian  $H_{Dirac}=vk_x\gamma^1+vk_y\gamma^2+m\gamma^0$ in a unit vector $\hat n(\vec k)=\frac{(vk_x,vk_y,m)}{\sqrt{v^2(k_x^2+k_y^2)+m^2}}$, the Berry curvature reads
 \begin{equation}
 \mathcal F(\vec k)=\frac{ \hat n\cdot d_{k_x}\hat n \times d_{k_y}\hat n}{2}.
 \end{equation}
 Pictorially the Berry phase accumulated is given by $1/2$ times the solid angle that $\hat n(\vec k)$ sweeps through during an adiabatic process. 
 
 To express the hall conductivity as one varies electron doping $x$, one first obtains the density of states near the Dirac cone as 
 \begin{eqnarray}
 \rho(E)=\frac{dN}{dE}
 =\frac{2\pi kdk }{v dk \frac{4\pi ^2 \sin \frac{\pi}{3}}{L_xL_y}},
 \end{eqnarray}
 where we used the area a state occupied in k-space for a triangular system with size $L_x,L_y$ as $\frac{4\pi ^2 \sin \frac{\pi}{3}}{L_xL_y}$ and $k$ as the norm of $\vec k$ since the dispersion is uniform around Dirac cones.
 At doping level of $x$,  the states in a circle around Dirac point with radius $k_c$ in valence band are emptied, $k_c$  given by
 \begin{eqnarray}
 \label{kc}
 \int_0^{k_c} dE \rho (E)=L_xL_y \frac{x}{4},\nonumber\\
 k_c^2=\pi x\sin \frac{\pi}{3} ,
 \end{eqnarray}
 where the factor $1/4$ comes from the $4$ Dirac cones in $2$ spinon conduction bands.
 
 The Berry curvature accumulated from states in a circle of radius $k_c$ is given by the solid angle swept by $\hat n$ from $|k|=0,\hat n=(0,0,1)$ to $|\vec k|=k_c,\hat n(\vec k)=\frac{(vk_x,vk_y,m)}{\sqrt{v^2k_c^2+m^2}}$, hence
 \begin{equation}
 \int_{|k|<k_c} d^2k \mathcal F(\vec k)= \frac{2\pi}{2} (1-\frac{m}{\sqrt{m^2+v^2 k_c^2}}),
 \end{equation}
 substituting $k_c$ from eq\eqref{kc} and $m,v$ from previous analysis, one gets the hall conductivity as the total Chern number from two spinon valence bands $C=2$, minus the contributions from states around Dirac cones as
 \begin{equation}
\sigma_{xy}=\frac{2e^2}{h}-\frac{4\pi e^2}{4\pi^2\hbar} (1-\frac{2\sqrt{3}\Delta}{\sqrt{12\Delta^2+4\sqrt{3}\pi x}}),\end{equation}
 where $\Delta=|\pi/6-\theta|$, which is the expression for $\sigma_{xy}$ as in eq\eqref{hall}.

 \begin{figure}
 \captionsetup{justification=raggedright}
    \centering
     \adjustbox{trim={.0\width} {.0\height} {.0\width} {.0\height},clip}
   { \includegraphics[width=0.95\textwidth]{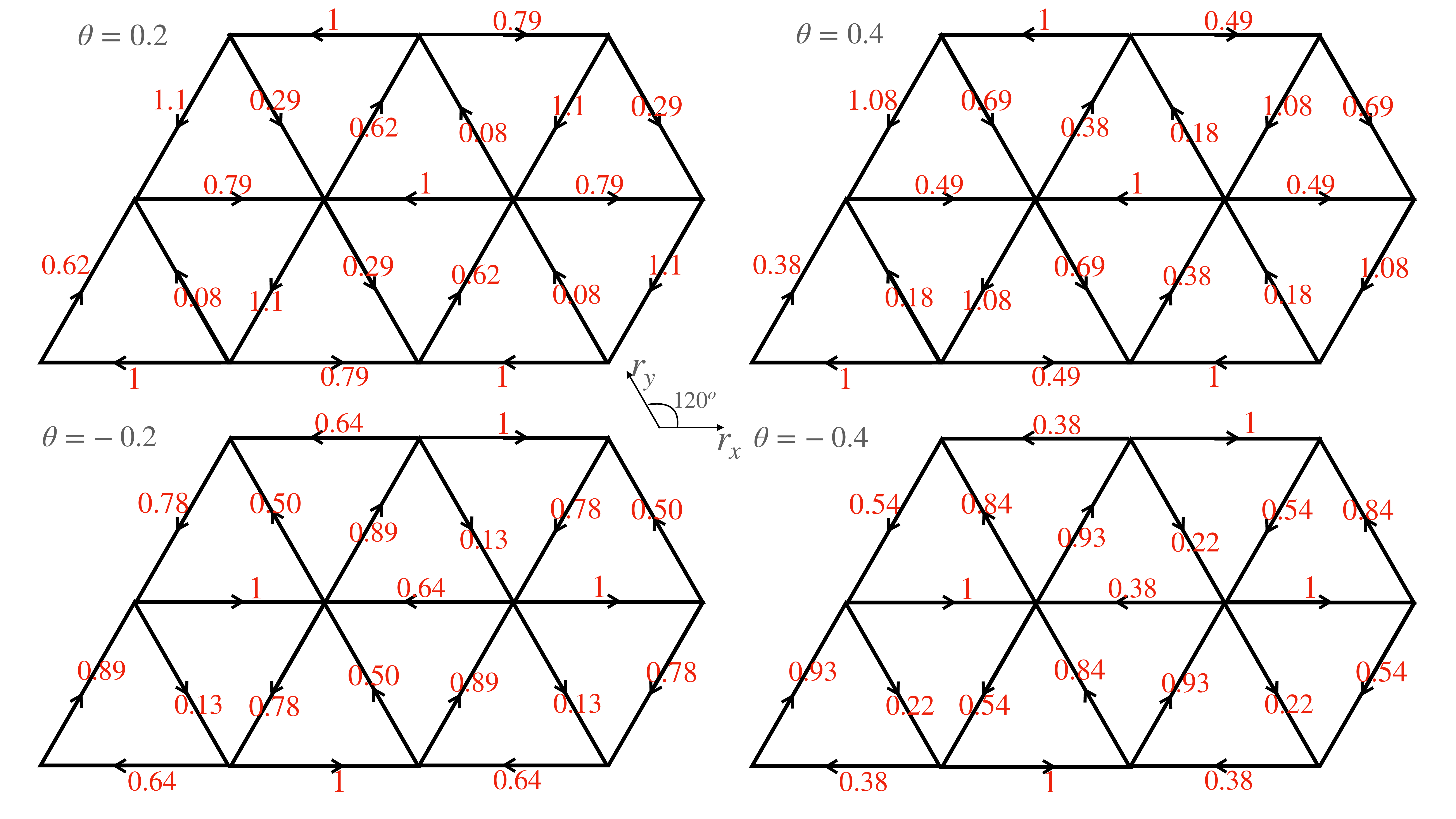}}
 \caption{The relative current strength $I_{bond\langle ij\rangle}=Im[c^\dagger_ic_j]$ of the chiral metal upon infinitesimal doping of holes shown as red numbers on the bond ($\theta$ as shown in upper left of each panel) with arrows indicating current directions. We assume the holons condense fully at $\Gamma$. The currents for infinitesimal electron doping are negative of those in hole doped case, and with a shift of the $y$ direction bonds.}
    \label{fig:chiral_metal_cur_sup}
\end{figure}

\section{Calculation of spectral function $A(k,\omega=0^+)$}
\label{app:spectral}
We define a plane-wave basis for electrons as
\begin{align}
c_k^\dagger=\frac{1}{L}\sum_r e^{ik\cdot r} c_i^\dagger,\nonumber\\
c_{k,A}^\dagger=\frac{1}{L}\sum_{r\in A} e^{ik\cdot r}c_i^\dagger,\nonumber\\
c_{k,B}^\dagger=\frac{1}{L}\sum_{r\in B} e^{ik\cdot r}c_i^\dagger.
\end{align}
The first one is used in defining the spectral function as one measures in arpes. The latter two correspond to plane waves on sublattices $A,B$, respectively. Taking the origin at $A$ site, these operators have the relation,
\begin{align}
c_k^\dagger=c_{k,A}^\dagger+c_{k,B}^\dagger,\nonumber\\
c_k^\dagger=c_{k+\vec G_1,A}^\dagger-c_{k+\vec G_1,B}^\dagger,
\end{align}
where we used $e^{i\vec G_1\cdot r_{AB}}=-1$, where $r_{AB}$ is the vector connecting sites on two sublattices.

The spectral function measured in the ground state at frequency $\omega=0^+$ is contributed by eigenstates $u_k^\dagger=\alpha_k c_{k, A}^\dagger+\beta_k c_{k,B}^\dagger$ on the Fermi pockets. The coefficients are related to those of the spinon eigenstates by holon condensed values,i.e.
\begin{align}
    \alpha_k=\sum_{k'}u_{b,k'} (A) \alpha_{k-k',f}\nonumber\\
    \beta_k=\sum_{k'}u_{b,k'} (B) \beta_{k-k',f},
\end{align}
where $\alpha_{k,f},\beta_{k,f}$ are the coefficients for sublattice $A,B$ wavefunctions for spinon eigenstates and $u_{b,k}(A,B)$ are the holon condensation discussed in section \ref{chiral_metal} eq \eqref{holonwfn}, non-vanishing only for $k=\Gamma,M_2$. Note that spinon states at $k,k+M_2$ have the same energy due to the projective symmetry $T_x$ with a momentum boost of $(0,\pi)$, i.e $M_2$ momentum.

This overlap of the plane wave basis for electrons reads
\begin{align}
\label{overlap}
\langle u_{k'}|c_k^\dagger|0\rangle=\delta_{k,k'}(\alpha_k+\beta_k)+\delta_{k,k'\pm\vec G_1}(\alpha_{k'}-\beta_{k'}).
\end{align}

The spectral function is given by 
\begin{align}
A(\vec k,\omega=0^+)=2\pi \int_{k'\in FS} dk' |\langle u_{k'}| c_k^\dagger |0\rangle|^2.
\end{align}
Since the original BZ measured in arpes is double the size of the reduced one, when measured momenta  $k$ lie outside of the reduced BZ, one effectively excites states inside the reduced BZ at its equivalent momentum (i.e., obtained by a translation of a reciprocal lattice vector $\vec G_1$).

Using eq \eqref{overlap} to express the spectral function, one finds a simple expression
\begin{align}
A(\vec k,\omega=0^+)=\begin{cases} 2\pi |\alpha_k+\beta_k|^2 & \vec k \in \textrm {reduced BZ}\\ 2\pi |\alpha_{ko}-\beta_{ko}|^2 & \vec {ko}=\vec k\pm \vec G_1 \in\textrm {reduced BZ}.\end{cases}
\end{align}
where $\vec {ko}$ is the equivalent momenta for $\vec k$ in the reduced BZ.

\section{A concrete mean field theory of the topological superconductor and deconfined transition}
\label{app:transition}
Here we try to construct a concrete mean field theory of the topological superconductor and writing down a mean field ansatz for it.

To simplify the construction of the bIQHE state, we introduce two slave bosons, each of which has filling $1/2$ per site.  We will hybridize them in the last stage so the final theory should be smoothly connect to that of one slave boson.   For the discussion of Mott transition at fixed density $n_c=1$, we use $U(1)$ slave boson theory, instead of the $SU(2)$ version.  The cost is that we can not keep track of the $C_6$ symmetry.  But we can still keep track of translations $T_x,T_y$.  We will show explicitly that our proposed ansatz is invariant under $T_x$ and $T_y$.

The modified slave boson theory is:

\begin{align}
    c_{i;\uparrow}&=b_i f_{i;\uparrow} \\
    c_{i;\downarrow}&=\tilde b_i  f_{i;\downarrow}
\end{align}

This will introduce two independent $U(1)$ gauge field: $a$ and $\tilde a$. But we will lock them together with a Higgs term $b_i^\dagger \tilde b_i+h.c.$.  Note the above parton construction does not satisfy the $SU(2)$ spin rotation symmetry.  We can only keep track of the $S_z$ quantum number. As a result, the state constructed in this way may break the  $SU(2)$ spin rotation down to $U(1)$.

Now the constraint is that:

\begin{align}
    n_{i;b}&=n_{i;f_{\uparrow}} \notag\\
    n_{i;\tilde b}&=n_{i;f_{\downarrow}}
\end{align}

So on average we have $n_b=n_{\tilde b}=\frac{1}{2}$ per site.  To describe the bIQHE phase, we do a further fractionalization:

\begin{align}
    b_i&=d_{i} \psi_{i;1}\notag\\
    \tilde b_i&=\tilde d_{i} \psi_{i;2} \notag\\
\end{align}
which introduces another two $U(1)$ gauge field $\gamma,\tilde \gamma$.

In total, the electron is fractionalized in the following way:

\begin{align}
\label{eoperator}
    c_{i;\uparrow}&=d_i \psi_{i;1}f_{i;\uparrow} \notag\\
    c_{i;\downarrow}&=\tilde d_{i}\psi_{i;2}f_{i;\downarrow}
\end{align}

We assign the charge in the following way: $\psi_{1}$ couples to $-a+A_c+\gamma$, $\psi_2$ couples to $-\tilde a+A_c+\tilde \gamma$.   $d$ couples to $-\gamma$ and $\tilde d$ couples to $-\tilde \gamma$. $f_{\uparrow}$ couples to $\frac{1}{2} A_s +a$ and $f_{\downarrow}$ couples to $-\frac{1}{2}A_s+\tilde a$.

We consider the follwing mean field ansatz:

\begin{align}
\label{partonmf}
    H=H_f+H_{d}+H_{\psi_1}+H_{\psi_2}-\lambda \sum_i (\psi^\dagger_{i;1} \psi_{i;2}+h.c.)
\end{align}

First, $H_f$ is just the mean field ansatz of the fermionic spinons.  $d,\tilde d$ hybridize together and have a mean field ansatz with  $0$ flux per unit cell ($T_x T_y=T_y T_x$.).   Their hybridization locks $\gamma=\tilde \gamma$, so we will only keep  $\gamma$ in the following analysis. We want the PSG of $d$ to be completely trivial. The non-trivial PSG of $b$ is inherited completely by $\psi$.   Because there are two orbitals formed by $d,\tilde d$, there can be a Chern insulator with $C_d=1$ even if the ansatz does not break translation symmetry. We show a generic mean-field for $d,\tilde d$ with such properties:
\begin{eqnarray}
H_d=\sum_{\langle ij\rangle} [d_{i}^\dagger d_{j}-\tilde d_{i}^\dagger \tilde d_{j}]+\sum_r p_1 d_{r}^\dagger (\tilde d_{r+\hat x}+e^{i\frac{2\pi}{3}}\tilde d_{r+\hat y}+e^{-i\frac{2\pi}{3}}\tilde d_{r-\hat x-\hat y})\nonumber\\
+\sum_r p_2 d_{r}^\dagger (\tilde d_{r-\hat x}+e^{i\frac{2\pi}{3}}\tilde d_{r-\hat y}+e^{-i\frac{2\pi}{3}}\tilde d_{r+\hat x+\hat y})+\textrm{h.c.},
\end{eqnarray}
which preserves translation and the mixed hopping between $d,\tilde d$ has angular momentum $2$. This mean-field dispersion is generically gapped except with particular parameters e.g. $p_1 p_2=0$ and may have non-vanishing Chern number, e.g. when $p_1=p_2=1+i$, the valence band has chern number $C_d=1$.

$\psi_1$ and $\psi_2$ have mean field ansatz with $T_x T_y=-T_y T_x$, which should be similar to that of spinon $f$ introduced in the main text.  They have Chern number $C_1=\pm 1$ and $C_2=1$. The mean-field we use for $\psi$ is those for the CSL1 in eq \eqref{hspinon} with $\theta=\pi/6\pm\delta,(\delta>0)$ for Chern number $C=\mp 1$, respectively.

In the following we analyze the cases with $\lambda=0$ and $\lambda \neq 0$ separately.  The topological superconductor discussed in the main text corresponds to $\lambda \neq 0$.  

\subsection{$\lambda=0$:  $p+ip$ superconductor}
We deal with the special case of $\lambda=0$ first. In this case, we need to deal with two $U(1)$ gauge field $a$ and $\tilde a$, which couples to $f_{\uparrow},b$ and $f_{\downarrow},\tilde b$ respectively.  First, integration of the spinon $f_\sigma$ gives:

\begin{equation}
    L_f=-\frac{1}{4\pi} \alpha_1 d \alpha_1 -\frac{1}{4\pi} \alpha_2 d \alpha_2 +\frac{1}{2\pi} (\frac{1}{2}A_s +a) d \alpha_1+\frac{1}{2\pi} (-\frac{1}{2}A_s +\tilde a) d \alpha_2
\end{equation}
where $\alpha_1,\alpha_2$ are introduced to describe the IQHE of the fermionic spinons.

Next we consider the ansatz with $C_d=-1,C_1=C_2=1$. The slave boson part has the action
\begin{equation}
    L_b=\frac{1}{4\pi}\beta d \beta -\frac{1}{2\pi} \gamma d \beta-\frac{1}{4\pi} \beta_1 d \beta_1-\frac{1}{4\pi} \beta_2 d \beta_2+\frac{1}{2\pi} (A_c+\gamma-a) d \beta_1+\frac{1}{2\pi}(A_c+\gamma-\tilde a) d \beta_2
\end{equation}
Integration of $\gamma$ locks $\beta=\beta_1+\beta_2$, then
\begin{equation}
    L_b=\frac{1}{4\pi}\beta_1 d \beta_2 +\frac{1}{4\pi}\beta_2 d \beta_1+\frac{1}{2\pi}(A_c-a) d \beta_1+\frac{1}{2\pi}(A_c -\tilde a) d \beta_2
\end{equation}

The final action is $L_c=L_b+L_f$. By integrating $a$ and $\tilde a$, we can lock $\beta_1=\alpha_1$ and $\beta_2=\alpha_2$. We can then reach the final action:

\begin{equation}
   L_c= -\frac{1}{4\pi} \alpha_1 d \alpha_1 -\frac{1}{4\pi} \alpha_2 d \alpha_2+\frac{1}{4\pi}\alpha_1 d \alpha_2+\frac{1}{4\pi} \alpha_2 d \alpha_1+\frac{1}{2\pi} (A_c+\frac{1}{2} A_s) d \alpha_1 +\frac{1}{2\pi} (A_c-\frac{1}{2} A_s ) d \alpha_2
\end{equation}

By relabeling $\alpha_1=\alpha_c+\alpha_s$ and $\alpha_2=\alpha_c-\alpha_s$, we get

\begin{eqnarray}
L_c=\frac{2}{2\pi} A_c d \alpha_c-\frac{4}{4\pi} \alpha_s d \alpha_s+\frac{1}{2\pi} A_s d \alpha_s
\end{eqnarray}

The above action is equivalent to the theory of  $p+ip$ superconductor, which is in the $\nu=2$ class of Kitaev's sixteen fold way classification\cite{kitaev_2006}.  $p+ip$ pairing is possible  in the spin-triplet channel of $c_{\uparrow} c_{\downarrow}$.  This is also an indication that the full $SU(2)$ spin rotation symmetry is broken in this state.
.
\subsection{$\lambda \neq 0$:  $d+id$ superconductor}

We will show that the case with $\lambda \neq 0$ gives a superconductor with the same topological property as the spin singlet $d+id$ superconductor.  

With $\lambda \neq 0$, the term $-\lambda \psi^\dagger_{i;1} \psi_{i;2}+h.c.$ locks $a=\tilde a$ (Note that we already lock $\gamma=\tilde \gamma$ by the hybridization between $d$ and $\tilde d$.).  After that, the action for the slave boson part should be modified as

\begin{equation}
    L_b=\frac{1}{4\pi}\beta_1 d \beta_2 +\frac{1}{4\pi}\beta_2 d \beta_1+\frac{1}{2\pi}(A_c-a) d \beta_1+\frac{1}{2\pi}(A_c -a) d \beta_2
\end{equation}
which is exactly the action in Eq.(19) of the main text.  

$L_f$ is also modified to
\begin{equation}
    L_f=-\frac{1}{4\pi} \alpha_1 d \alpha_1 -\frac{1}{4\pi} \alpha_2 d \alpha_2 +\frac{1}{2\pi} (\frac{1}{2}A_s +a) d \alpha_1+\frac{1}{2\pi} (-\frac{1}{2}A_s +a) d \alpha_2
\end{equation}
which is the same as Eq.(18) in the main text.

Then the analysis in the main text shows that this state is topologically equivalent to $d+id$ superconductor.

Finally, let us comment on the spin rotation symmetry.  Our formalization only explicitly preserve $U(1)$ spin rotation corresponding to $S_z$ because we have two different slave bosons $b,\tilde b$ for the two spins.  When $\lambda=0$, there are two $U(1)$ gauge fields $a,\tilde a$. This strongly breaks spin rotation symmetry, and thus we get a state consistent with a spin-triplet  $p+ip$ superconductor. When we increase $\lambda$, we couple $b_i$ and $\tilde b_i$ together. In the limit that this hybridization is very large, we can keep only one $b'_i=\frac{1}{\sqrt{2}}(b_i+\tilde b_i)$ and then we recover the usual slave boson theory, which explicitly preserve the $SU(2)$ spin rotation.  Therefore, we believe the phase constructed in this section can smoothly crossover to the spin-singlet $d+id$ superconductor when we increase the hybridization $\lambda$.

\subsection{Model wavefunction\label{appendix:model_wavefunction}}

A wavefunction can be written down based on the above parton construction:
\begin{align}
    \Psi(\vec x_1,\cdots \vec x_N)=P_G \Psi_{d,\tilde d}(\vec x_1,\cdots \vec x_N)\Psi_{\psi}(\vec x_1,\cdots \vec x_N)\Psi_f(\vec x_1,\cdots \vec x_N),
\end{align}
where $P_G$ is the projection to enforce the constraints in eq \eqref{eoperator} and $\Psi_{d,\tilde d},\Psi_\psi, \Psi_f$ are the ground state wave functions of the mean field discussed in eq \eqref{partonmf}.

\subsection{Deconfined CSL to superconductor transition}

The critical point is described by changing $C_1$ from $-1$ to $1$.  This is captured by mass changing of two Dirac cones $\chi_{1,2}$ with momenta $(\pi/3,\pi/3),(\pi//3,-2\pi/3)$, obtained by the changing of $\theta$ in $\psi_1$ mean field eq \eqref{hspinon} from $\pi/6+\delta$ to $\pi/6-\delta,(\delta>0)$.  Because $\psi_1$ satisfies the projective translation symmetry $T_xT_y=-T_y T_x$, the mass of the two Dirac cones $\chi_{1,2}$ at $\theta=\pi/6$ are guaranteed to be the same by the translation symmetry.  The translation acts projectively on two Dirac cones as
\begin{align}
    T_x:(\chi_{1},\chi_2)^T\rightarrow e^{-i\frac{\pi}{3}} (\eta^x\chi_{2},\eta^x\chi_1)^T,\nonumber\\
    T_y: (\chi_{1},\chi_2)^T\rightarrow diag(e^{-i\frac{\pi}{3}},-e^{-i\frac{\pi}{3}})(\chi_{1},\chi_2)^T,
\end{align}
where $\eta^x$ is the Pauli matrix on the Dirac indices of the Dirac fermions.

Adding up the Dirac fermions and actions for $\psi_2,d,\tilde d,f$ etc, one has for the critical action,(identifying $\gamma=\tilde \gamma,a=\tilde a$)
\begin{align}
\label{app:cri}
    \mathcal L_{cri}=\sum _{i=1,2} \bar\chi_i [\eta_\mu (i\partial_\mu-a_\mu+A_{c,\mu}+\gamma_\mu)\chi_i+m_i\bar \chi_i\chi_i\nonumber\\
    +\frac{1}{4\pi}(\beta d\beta-2\gamma d\beta-\beta_2d\beta_2+2(A_c+\gamma-a)d\beta_2)\nonumber\\
    -\frac{1}{4\pi} \alpha_1 d \alpha_1 -\frac{1}{4\pi} \alpha_2 d \alpha_2 +\frac{1}{2\pi} (\frac{1}{2}A_s +a) d \alpha_1+\frac{1}{2\pi} (-\frac{1}{2}A_s +a) d \alpha_2,
\end{align}
where the first line describes critical theory for $\psi_1$ ($m=0$ at critical point) , the second line describes $d,\tilde d,\psi_2$, and last line for the spinon action.

With nonzero masses ($m_1=m_2$ by the projective translation symmetry), one integrates out $\chi$. To be more precise, we should introduce another field $\beta_1$ for describing the state $C_1=\pm 1$ after integrating out $\chi$: $\frac{\mp 1}{4\pi} \beta_1 d \beta_1 +\frac{1}{2\pi} \beta_1 d(A_c+\gamma-a)$. 

When $m_1=m_2>0$, $\psi_1$ fills a band with $C_1=1$ and the theory should describe a $d+id$ superconductor.Rewriting $\gamma=\Gamma+a$,$\alpha_c=\frac{\alpha_1+\alpha_2}{2},\alpha_s=\frac{\alpha_1-\alpha_2}{2}$ and integrating out $a$ gives $\beta=2\alpha_c$. The critical action becomes
\begin{align}
     \mathcal L_{cri}=\sum_{i=1,2} \bar\chi_i [\eta_\mu (i\partial_\mu+A_{c,\mu}+\Gamma_\mu)\chi_i+m_i\bar \chi_i\chi_i\nonumber\\
    +\frac{1}{4\pi}(2\alpha_c d\alpha_c-4\Gamma d\alpha_c-\beta_2d\beta_2+2(A_c+\Gamma)d\beta_2)\nonumber\\-\frac{2}{4\pi} \alpha_s d \alpha_s +\frac{1}{2\pi} A_s d \alpha_s.
\end{align}
 Integrating out $\chi_i$, $\Gamma$ gives $2\alpha_c=\beta_1+\beta_2$ and an action
\begin{align}
    \mathcal L_{d+id}=\frac{1}{4\pi}(-2\alpha_cd\alpha_c+4\alpha_cd\beta_2-2\beta_2d\beta_2+4A_cd\alpha_c)\nonumber\\
   -\frac{2}{4\pi} \alpha_s d \alpha_s +\frac{1}{2\pi} A_s d \alpha_s .
\end{align}
If one identifies $\tilde \beta=\alpha_c-\beta_2$ and rewriting the action in $\tilde \beta, \alpha_c$, the action is identical to the $d+id$ superconductor action in eq \eqref{dwavesc} with gapless $\alpha_c$ mode coupled to $A_c$, hence superconducting.

When $m_1=m_2<0$,$\psi_1$ fills a band with $C_1=-1$ and the theory should describe a chiral spin liquid (CSL1). Integrating out $\chi_i$ gives $\frac{1}{4\pi} \beta_1 d \beta_1 +\frac{1}{2\pi} \beta_1 d(A_c+\gamma-a)$, and integrating out $\gamma$ in eq \eqref{app:cri} gives $\beta=-(\beta_1+\beta_2)$ and an action
\begin{align}
    \mathcal L_{CSL}=\frac{1}{4\pi}[2\beta_1d\beta_1+2\beta_1d\beta_2+2(A_c-a)d(\beta_1+\beta_2)]\nonumber\\
   -\frac{1}{4\pi} \alpha_1 d \alpha_1+\frac{1}{2\pi} (a+\frac{1}{2}A_s) d \alpha_1
   -\frac{1}{4\pi} \alpha_2 d \alpha_2 +\frac{1}{2\pi} (a-\frac{1}{2}A_s) d \alpha_2.
\end{align}
Integrating $\beta_2$ locks $\beta_1=a-A_c$. Substituting it into the action, we find that the terms from the slave boson part cancel each other and we are left with the action from the fermionic spinon:

\begin{align}
    \mathcal L_{CSL}=
   -\frac{1}{4\pi} \alpha_1 d \alpha_1+\frac{1}{2\pi} (a+\frac{1}{2}A_s) d \alpha_1
   -\frac{1}{4\pi} \alpha_2 d \alpha_2 +\frac{1}{2\pi} (a-\frac{1}{2}A_s) d \alpha_2. 
\end{align}

Integrating $a$ locks $\alpha_1=-\alpha_2=\alpha$, and we obtain the final action for the $U(1)_2$ CSL1:

\begin{align}
    \mathcal L_{CSL}=
   -\frac{2}{4\pi} \alpha d \alpha +\frac{1}{2\pi} A_s d \alpha.
\end{align}

\end{widetext}

\bibliography{temp_doped}

\end{document}